\documentclass[secnumarabic,twocolumn,nofootinbib,tightenlines,aps,superscriptaddress]{revtex4}
\usepackage{amsfonts}
\usepackage{amsmath}
\usepackage{amssymb}
\usepackage{graphicx}%
\usepackage{bm}%
\usepackage{setspace}
\usepackage{longtable}\makeatletter
\let\@currsize\normalsize
\makeatother

\begin{document}

\title{ Nonlinear Dynamics of Ion Concentration Polarization in Porous Media: 
The Leaky Membrane Model }

\author{E. Victoria Dydek}
\affiliation{Department of Chemical Engineering,\\Massachusetts Institute of Technology, Cambridge, MA 02139 USA}
\author{Martin Z. Bazant }
\affiliation{Department of Chemical Engineering,\\Massachusetts Institute of Technology, Cambridge, MA 02139 USA}
\affiliation{Department of Mathematics,\\Massachusetts Institute of Technology, Cambridge, MA 02139 USA}
\date{\today}

\begin{abstract}
The conductivity of highly charged membranes is nearly constant, due to counter-ions screening pore surfaces.  Weakly charged porous media, or ``leaky membranes", also contain a significant concentration of co-ions, whose depletion at high current leads to ion concentration polarization and conductivity shock waves.  To describe these nonlinear phenomena the absence of electro-osmotic flow, a simple Leaky Membrane Model is formulated, based on macroscopic electroneutrality and Nernst-Planck ionic fluxes. The model is solved in cases of unsupported binary electrolytes: steady conduction from a reservoir to a cation-selective surface, transient response to a current step, steady conduction to a flow-through porous electrode, and steady conduction between cation-selective surfaces in cross flow. The last problem is motivated by separations in leaky membranes, such as shock electrodialysis.  The article begins with a tribute to Neal Amundson, whose pioneering work on shock waves in chromatography involved similar mathematics.\end{abstract}

\maketitle

\begin{center}
{\it Dedication by Martin Z. Bazant }
\end{center}

This article is dedicated to the memory of Neal R. Amundson, the ``father of modern chemical engineering"~\cite{bates2011}, who  brought mathematical rigor to the fields of transport phenomena and reactor engineering~\cite{aris_book,rhee_book}.  His education, teaching and research were truly interdisciplinary, long before that term came into fashion.  His early education (BS 1937, MS 1941) was in  Chemical Engineering, the field of his primary faculty appointments at University of Minnesota (1949-1977) and University of Houston (1977-2011) and lifelong professional focus.  He is famous for leading the Department of Chemical Engineering at Minnesota to lasting national prominence, as its chair for 25 years --  starting at age 33, only two years after being hired.  It is perhaps surprising then, that his most advanced degree at Minnesota (PhD 1945) and his early teaching as an Assistant Professor (1945-1947) were not in Chemical Engineering, but in Mathematics.  

Amundson revolutionized the way that chemical engineers design systems for separations, heat transfer and adsorption, by replacing empirical principles with rigorous models based on partial differential equations (PDE). His PhD thesis (1945)~\cite{amundson_thesis} and early papers (1948-1952)~\cite{amundson1948,amundson1950,lapidus1952} involved analytical solutions of PDEs for flow-through adsorption in porous media, which are still used today in chromatography, electrophoresis and ion exchange.  His work built upon the seminal paper of Thomas (1944)~\cite{thomas1944} and preceded those of Goldstein (1953)~\cite{goldstein1953_1,goldstein1953_2}, which are better known in applied mathematics~\cite{whitham_book}.  

A major achievement of Amundson was to show that flow-through adsorption processes described by the PDE,
\begin{equation}
\frac{\partial f(c) } {\partial t} + v \frac{\partial c}{\partial x} = D \frac{\partial^2 c}{\partial x^2}, \ \ c(0,t) = c_0(t)   \label{eq:ads}
\end{equation}
where $c$ is the flowing concentration and $f(c)$ is total (adsorbed $+$ flowing) concentration per volume in local equilibrium,
lead to nonlinear kinematic waves~\cite{whitham_book,rhee_book}. Neglecting diffusion ($D=0$), the model reduces to a first-order quasi-linear PDE, which can be solved in implicit form, 
\begin{equation}
c = c_0\left( t- \frac{x \, f^\prime(c)}{v} \right),
\end{equation}
by Lagrange's Method of Characteristics, as explained in a series of papers by Rhee, Aris and Amundson (1970-1972)~\cite{rhee1970,rhee1971,rhee1972}.  For most boundary and initial conditions, kinematic waves eventually ``break" and produce shock waves, or propagating discontinuities, which are sharp in the limit $D=0$.   Lapidus and Amundson (1952)~\cite{lapidus1952}  showed that these discontinuities are broadened by diffusion ($D>0$) around the same time that Hopf~\cite{hopf1950} and Cole~\cite{cole1951} famously solved Burgers equation in fluid mechanics~\cite{whitham_book}.  Amundson, Aris and Swanson (1965)~\cite{amundson1965} rigorously explained and analyzed  the sharp concentration bands  arising in chromatography and ion exchange as concentration shocks in porous exchange beds~\cite{rhee1971,rhee1972,davis1987,rhee_book}.  

Upon being invited to contribute to this special issue, I set out to learn more about the man and his life's work.  At first, I was struck by the unusual parallels with my own career, since I too began graduate study and teaching in mathematics before moving to  chemical engineering and holding a joint appointment. What impressed me most, however, were the  parallels in research.  Without knowing Admunson's seminal work on exchange beds, A. Mani and I recently developed a theory of ``deionization shocks" in porous media~\cite{mani2011desalination} that bears intriguing mathematical similarities;  surface conduction within the pores and bulk ionic current play the roles of surface adsorption and pressure-driven flow, respectively.  This phenomenon was first discovered in microfluidics by Mani, Zangle and Santiago (2009)~\cite{mani2009propagation,zangle2009propagation,zangle2010theory,zangle2010constantvoltage}, but its extension to porous media -- analogous to  Amundson's work -- is more general (e.g. decoupling the directions of flow and current) and paves the way for practical applications.  Indeed,  the first experiment in my new laboratory in the Department of Chemical Engineering~\cite{deng2013} applies the shock theory to water deionization by ``shock electrodialysis"~\cite{shockED_patent} and relies on mathematical analysis for finite-length pores~\cite{dydek2011overlimiting}  to interpret the data, as elaborated in this article.

Since Amundson and I both began our careers in mathematics, it is perhaps not surprising that we ended up doing similar research in chemical engineering.  At MIT, I used  to teach 18.311 Principles of Applied Mathematics, which introduced PDEs to undergraduates, starting with first-order quasilinear equations (applied to traffic flow)~\cite{haberman_book}. In contrast, chemical engineering courses up to the graduate level (such as 10.50 Analysis of Transport Phenomena, which I teach with W. Deen~\cite{deen_book} following Amundson's co-teaching model~\cite{bates2011}) mainly cover the solution of linear parabolic PDEs, including the Finite Fourier Transform method championed by Amundson~\cite{aris_book}. As a result, most chemical engineering students today do not know the Method of Characteristics, even though it is the mathematical basis for theories of chromatography ~\cite{rhee1970,rhee_book}, gas dynamics~\cite{bradley_book,whitham_book}, electrokinetic soil remediation~\cite{probstein1993,shapiro1993,jacobs1994,kamran2012}, capillary electrophoresis~\cite{ghosal2010,ghosal2012,chen2012a,chen2012b,chen2012c}, and ion  concentration polarization in microchannels~\cite{mani2009propagation,zangle2009propagation,zangle2010theory} and porous media~\cite{mani2011desalination,yaroshchuk2012acis} (the focus of this article).  

Perhaps this special issue of {\it AIChE Journal} can serve as a call to reinvigorate the teaching of mathematical methods introduced by Amundson to our field, which provide physical insights and useful formulae, too often overlooked in the Computer Age.   $\Box$

\vspace{0.1in}

%\clearpage

\section*{Introduction}

When current is passed through an electrolyte to an ion selective surface (such as an ion-exchange membrane, micro/nanochannel junction or electrode), the passage of certain ions, and the rejection of others, generally lead to concentration variations and voltage losses (or internal resistance), known as ``ion concentration polarization" (ICP). Under classical assumptions of electroneutrality without convection or homogeneous reactions, the current is limited by electrodiffusion, when the concentration of the active species vanishes at the selective surface~\cite{probstein1994}.  In a neutral binary electrolyte, the current appears to be limited by diffusion alone, since the concentration profiles evolve according to a pure diffusion equation, but electromigration and diffusion conspire to determine the effective (ambipolar) diffusivity~\cite{newman_book}.  Both species diffuse in the same direction, but electromigration enhances the flux of the active species and opposes the flux of the inactive species (and cancels it in steady state).

Despite this theoretical speed limit, overlimiting current (OLC) has been observed in a  variety of systems involving membranes, porous media, and micro/nanochannels. Elucidating mechanisms for OLC remains a central question in membrane science and chemical engineering~\cite{nikonenko2010intensive}.  In free electrolyte solutions, there are two fundamental mechanisms for OLC -- chemical and physical -- each of which affects ICP in different ways.  

Chemical mechanisms for OLC involve the production of additional ions (from solvent decomposition, H$^+$ and OH$^-$ in water) and/or the loss of surface selectivity (from charge regulation of a membrane or nanochannel or from side reactions at an electrode) in order to reduce ICP and maintain ionic conductivity at high currents~\cite{belova2006,pismen2007,nikonenko2010intensive}.  Andersen et al. (2012)~\cite{andersen2012} recently showed that both phenomena are needed to achieve significant OLC via the phenomenon of ``current-induced membrane discharge" (CIMD). In particular, for aqueous systems, CIMD can result from bulk water splitting  coupled to charge regulation of the membrane, e.g. by proton adsorption in anion exchange membranes. 

Physical mechanisms for OLC involve the amplification of a different transport mechanism, which allows ions to reach the selective surface faster than by quasi-neutral electrodiffusion in the region of strongest ICP near the selective surface.  The best-known example is the Rubinstein-Zaltzman electro-osmotic instability~\cite{zaltzman2007electro,rubinstein2000,rubinstein2001,rubinstein2005}, which has recently been observed in micro/nanofluidic systems\cite{rubinstein2008direct,yossifon2008,chang2012}.
In porous media and microchannels, the presence of charged double layers on the side walls, aligned with the direction of current, allows OLC to be sustained by the additional transport mechanisms of surface conduction (SC)~\cite{dydek2011overlimiting} and electro-osmotic flow (EOF)~\cite{yaroshchuk2011coupled,dydek2011overlimiting}.  At lower (under-limiting) currents, ICP has also been observed experimentally in an electroosmotic pump with a porous glass frit\cite{suss2011electroosmotic} and in porous electrodes in a system designed for capacitive desalination\cite{suss2012capacitive}.

Advances in microfluidics over the past ten years have enabled the direct observation of ICP during OLC. Steady, sharply defined depletion zones, sometimes containing internal electro-osmotic vortices, have been observed  near the junctions of microchannels and nanochannels by by J. Han's group since 2005 \cite{wang2005million,kim2010direct,kim2013} and have been shown to be affected by the microchannel geometry \cite{kim2007concentration,yossifon2010controlling}.  Of particular note for this work, Mani, Zangle and Santiago (2009) have shown that in very thin (1 $\mu$m) channels, these depletion interfaces can propagate as shock waves under constant current \cite{mani2009propagation,zangle2009propagation,mani2011desalination}. Zangle et al. (2010)\cite{zangle2010theory} give an insightful review of these experiments and the theory behind them, which is mathematically similar to Amundson's theory of chromatography, as noted above. 

While the original experiments and theory were limited to single microchannels,  Mani and Bazant (2011)\cite{mani2011desalination} predicted the possibility of propagating deionization shocks in porous media and formulated general nonlinear PDEs to describe volume-averaged ICP at the macroscopic scale, driven by SC within charged nanopores, neglecting EOF as a first approximation. They obtained analytical similarity solutions for power-law variations in microstructure and analyzed the internal structure and dynamical stability of deionization shocks in three dimensions.    The formal volume averaging of their macroscopic ICP model is analogous to Amundson's theory of surface adsorption in fixed beds~\cite{amundson1948,amundson1950,lapidus1952,amundson1965,rhee1970} and Helfferich's early models of ion exchange membranes~\cite{helfferich_book}, but the PDEs are solved under general conditions of strong ICP with diffusion. Although porous ICP equations also provide simple area-averaged descriptions of  microfluidic devices, they are also more general because the flow and current directions can be decoupled in three dimensions, opening some new possibilities for separations.

In this article, we develop a fundamental picture of ICP dynamics in finite-size porous domains, including effects of simultaneous pressure-driven fluid flow and applied current, leading to two-dimensional concentration variations. The analysis is based on the PDEs of Mani and Bazant~\cite{mani2011desalination} for nanopores (or nanochannels) in the SC regime with forced convection, neglecting electro-osmotic flows that dominate in larger pores at the micron scale~\cite{dydek2011overlimiting}.  Borrowing a term of  A. Yaroshchuk~\cite{yaroshchuk2012nano},  we will refer to this as the ``Leaky Membrane Model".  All of the example calculations here are motivated by experiments in our group aimed at establishing surface-transport mechanisms for OLC and harnessing ICP dynamics in porous media for novel separations~\cite{deng2013}.  

Before we begin, we would like to draw attention to the recent work of Yaroshchuk, connecting these ideas to classical membrane science~\cite{helfferich_book} via in a theory of ``leaky membranes" and performing some similar transient~\cite{yaroshchuk2012acis} and steady state~\cite{yaroshchuk2012cp,yaroshchuk2012nano} calculations in one dimension, without flow.  His analysis is based on ``virtual concentrations" in local thermodynamic equilibrium  with a hypothetical ionic reservoir across each thin slice of the porous medium. The results can be left in general form or connected to specific quasi-equilibrium local models, such as the Poisson-Boltzmann model with fixed surface charge in a straight channel, with thin or thick double layers~\cite{yaroshchuk2011acis}.   This is an analytical limit of the full model of Mani et al~\cite{mani2009propagation} for thick double layers with effective longitudinal transport coefficients obtained numerically by integrating over the cross section. Our approach uses the physical volume-averaged concentration variables (defined in the next section) and the slowly varying, macroscopic part of the potential of mean force.  As such, interfacial voltages at the ends of the porous domain must be added to describe experimental data, but this can be done accurately by modeling or measuring the electrochemical series resistances and open circuit voltage~\cite{deng2013}.

\begin{figure*}
\begin{center}
\includegraphics[width=6.5in]{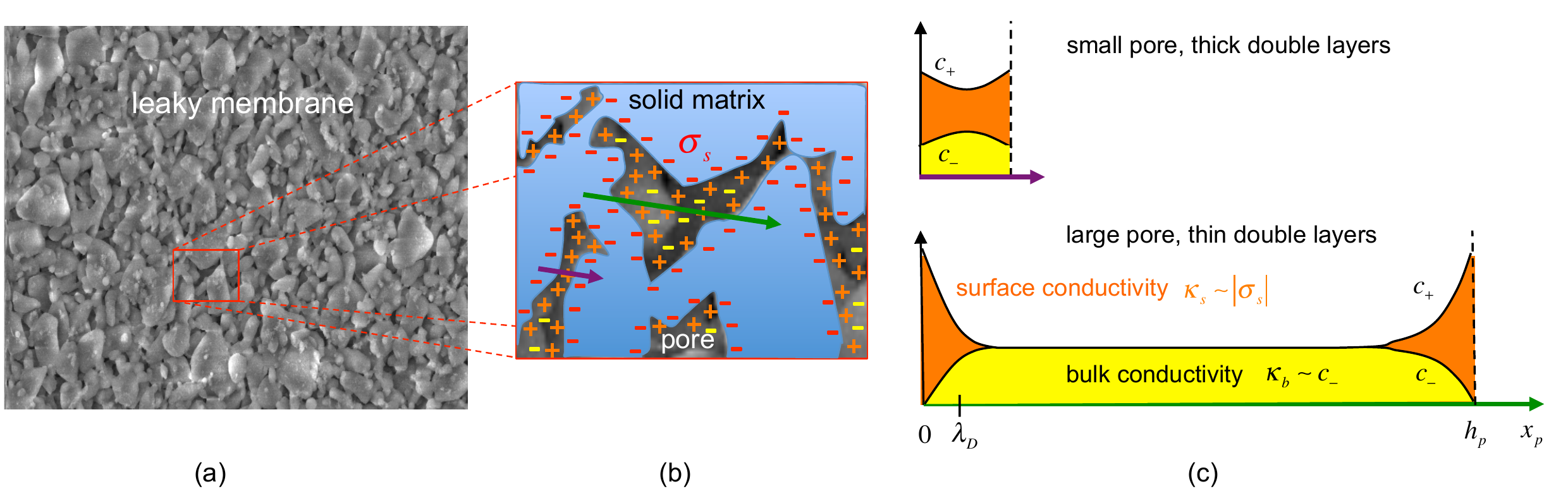}
\caption{ Physical picture of a leaky membrane.  (a) SEM image of a silica glass frit with 557 nm mean pore size, which can sustain over-limiting current and deionization shocks  (Deng et al.~\cite{deng2013}.) (b) Fixed surface charges (red) of density $\sigma_s$ per pore area and $\rho_s$ per total volume, and mobile counter-ions (orange) and co-ions (yellow) in the pores of concentration $c_\pm$ per total volume. (c) Sketch of the quasi-equilibrium ion profiles in small and large pores, relative to the Debye length $\lambda_D$. The surface conductivity scales with the total screening charge (orange areas), and the bulk, depletable conductivity scales with the co-ion charge (yellow areas). \label{fig:leaky} }\end{center}
\end{figure*}

\section*{ Leaky Membrane Model  }

\subsection*{ {\it General Formulation } }

Consider a charged porous medium of porosity $\epsilon_p$, internal pore surface area/volume $a_p$, pore surface charge/area $\sigma_s$ filled with an electrolyte, containing ions of charge $z_ie$ and concentration $c_i$ (per pore volume). 
An example of a porous silica glass frit is shown in Fig.~\ref{fig:leaky}.
Charge conservation implies,
\begin{equation}
\epsilon_p \sum_i z_i e c_i + a_p \sigma_s = 0  
\end{equation}
which can be written as a balance of surface charge per pore volume,
\begin{equation}
\rho_s =  \frac{a_p \sigma_s}{\epsilon_p} = - \sum_i z_i e c_i,  \label{eq:neut}
\end{equation}
with the electrolyte charge density.

The electroneutrality condition (\ref{eq:neut}) determines the relative importance of surface charge.
Let $c_0$ be a typical concentration of co-ions (of the same sign as the surface charge), which sets the scale for neutral salt permeating the porous medium. There are two limiting cases, illustrated in Fig.~\ref{fig:leaky}(c). In the membrane limit, $|\rho_s| \gg e c_0$, co-ions are strongly excluded, and the porous medium maintains a large, constant conductivity from nearly uniformly distributed counter-ions (of the opposite sign as the surface charge). In the opposite limit, $|\rho_s| \ll e c_0$, naively one would expect classical electrodiffusion of a quasi-neutral electrolyte, only with diffusivities rescaled to account for the tortuosity of the medium. This is indeed the case at low currents, but it turns out that the surface charge is a singular perturbation.  As we shall see, a ``leaky membrane" with $|\rho_s| > 0$, no matter how small, is fundamentally different from an uncharged porous medium with $\rho_s=0$.

The simplest approximation for the nonlinear dynamics of the electrolyte in a charged porous medium, which we call the Leaky Membrane Model (LMM), combines the macroscopic electroneutrality condition (\ref{eq:neut}) with homogenized Nernst-Planck equations~\cite{mani2011desalination}
\begin{equation}
\epsilon_p \frac{ \partial c_i }{\partial t} +  \epsilon_p\nabla \cdot \mathbf{F}_i = R_i  \label{eq:NP} 
\end{equation}
\begin{equation}
\mathbf{F}_i = \mathbf{u} c_i -  D_i \left( \nabla c_i + \frac{ z_i e c_i}{k_B T} \nabla \phi \right)
\end{equation}
where $\mathbf{F}_i$ is the macroscopic flux density  of species $i$ (per cross-sectional pore area), $D_i$ is the macroscopic chemical diffusivity  (with tortuosity correction~\cite{ferguson2012}), $\phi$ is the slowly varying, non-equilibrium part of the electrostatic potential of mean force, $\mathbf{u}$ is the mean fluid velocity, and $R_i$ is the mean reaction rate (per volume) producing species $i$. The current density  is 
\begin{equation}
\mathbf{J} = \sum_i z_i e \mathbf{F}_i
\end{equation}
(per cross-sectional pore area). 
For binary electrolytes, the LMM can also be cast in terms of bulk and surface conductivities (defined below)~\cite{mani2011desalination}.  

In a concentrated solution, the chemical diffusivity depends on the ionic concentrations~\cite{newman_book,ACR2013},
\begin{equation}
D_i = D_i^0 \left( 1 + \frac{\partial \ln \gamma_i}{\partial \ln c_i} \right)   \label{eq:Dconc}
\end{equation}
where $D_i^0$ is the tracer diffusivity in a dilute solution (which also generally depends on concentration~\cite{ACR2013,ferguson2012,nauman2001}) and $\gamma_i$ is the molar activity coefficient. Using  (\ref{eq:Dconc}) in (\ref{eq:NP}) leads to ``modified Poisson-Nernst-Planck equations"~\cite{kilic2007b,lai2011a}, which are used to account for thermodynamic effects in the nonlinear dynamics of electrolytes~\cite{large_acis,olesen2010}. In a general formulation based on non-equilibrium thermodynamics~\cite{ACR2013}, the ionic fluxes and reaction rate are related to electrochemical potentials, defined as variational derivatives of the total free energy functional. For systems with phase transformations (such as precipitation or phase separation), the Nernst-Planck equations (\ref{eq:NP}) become generalized Cahn-Hilliard-reaction models~\cite{ACR2013} coupled with macroscopic quasi-neutrality (\ref{eq:neut}) in the LMM framework. Coupled diffusive fluxes, e.g. friction between different species in a  Stefan-Maxwell formulation, can also be important for large ions or charge colloids in porous media~\cite{biesheuvel2011twofluid}.

Here we focus mostly on dilute solutions, where $D_i$ is constant or simply a function of the local salt concentration. 

\subsection*{ {\it Theoretical Justification }}

In the absence of flow and reactions, the LMM can be derived from the microscopic Poisson-Nernst-Planck  (PNP) equations within the pores by taking the limit of thin double layers and area averaging~\cite{mani2009propagation}, by formal  homogenization of the microscopic PDEs for arbitrary double layer thickness~\cite{schmuck_preprint}, and by assuming local thermodynamic quasi-equilibrium at the pore scale~\cite{yaroshchuk2012acis}. The microscopic potential of mean force within the pores is $\phi + \psi_{eq}$, where $\phi$ is the slowly varying part reflecting macroscopic departures from equilibrium and $\psi_{eq}$ is the rapidly varying correction due to quasi-equilibrium local interactions between the ions and the surface charge, constrained by the slowly varying mean ion concentrations $c_i$.  In the microscopic PNP equations, the surface charge per internal pore area, $\sigma_s$, enters via the electrostatic boundary condition on the pore walls, but after homogenization the macroscopic Nernst-Planck equations (\ref{eq:NP}) are simply augmented by a quasi-neutrality condition (\ref{eq:neut}) that includes the surface charge per volume $\rho_s$.   Effectively, the local quasi equilibrium charge fluctuations associated with $\psi_{eq}$ are ``integrated out" by homogenization to macroscopic length scales in the porous medium.

Fluid flow in charged porous media is much more complicated to homogenize rigorously, and no simple approximations are currently available. The difficulty is that flows within the pores are strongly coupled to the ion profiles via (locally linear) electrokinetic phenomena and lead to complex dispersion effects, modifying $D_i$ by nonuniform convection in the porous medium. Classical Taylor dispersion~\cite{yaroshchuk2011coupled} is often negligible compared to internal electro-osmotic convection~\cite{dydek2011overlimiting} and eddy dispersion~\cite{deng2013}.  In the simplest version of the LMM considered here, we neglect electroconvection and dispersion in (\ref{eq:NP}) and simply assume an imposed pressure-driven flow. 

In this work, we also neglect the reaction rate $R_i$, but reactions are important in many situations, such as electrokinetic remediation~\cite{probstein1993,shapiro1993,jacobs1994,kamran2012} and  porous electrode charging~\cite{biesheuvel2011,biesheuvel2012,ferguson2012}, and provide an additional source of nonlinearity.  In particular, the surface charge density, $\sigma_s$, is generally a function of the local electrolyte composition via the specific adsorption of ions. This phenomenon of ``charge regulation" is crucial for the quantitative interpretation of shock electrodialysis experiments using LMM~\cite{deng2013} and also underlies the theory of current-induced membrane discharge~\cite{andersen2012}. Assuming fast adsorption kinetics, the LMM then closely resembles Amundson's classical models of exchange beds of the form (\ref{eq:ads}), except that the LMM also accounts for the electrokinetic coupling between ion adsorption via the macroscopic charge balance (\ref{eq:neut}).  

\subsection*{ {\it Uniform Membrane Charge in a Binary Electrolyte }}

In order to highlight the nonlinear dynamics of ion transport in porous media, we adopt the simplest form of the LMM.   As noted above, we neglect reactions ($R_i=0$) and charge regulation ($\frac{\partial \rho_s}{\partial t}=0$). We also impose a pressure-driven Darcy flow $\mathbf{u}$ without accounting for dispersion or electrokinetic phenomena. In particular, we consider only the representative cases of uniform flow, either parallel or perpendicular to the applied current. We further assume a uniform porous medium with constant microstructure and charge ($\rho_s,\epsilon_p=$constants). For ease of calculations, we also make the standard theoretical assumption of a symmetric binary $z:z$ electrolyte with equal ionic diffusivities $D$ (including the tortuosity correction).  See Mani and Bazant~\cite{mani2011desalination} for extensions to asymmetric binary electrolytes and nonuniform porous media (with uniform flow) and Andersen et al.~\cite{andersen2012} for the full LMM (without flow) for a multicomponent (four species) electrolyte in a leaky membrane  with charge regulation (proton adsorption) and homogeneous reactions (water self-ionization).

With these assumptions,  the LMM takes the form
\begin{eqnarray}
\frac{ \partial \tilde{c}_\pm }{\partial \tilde{t}} &=& - \tilde{\nabla}\cdot\tilde{\mathbf{F}}_\pm   
		\label{eq:bin1} \\
\tilde{\mathbf{F}}_\pm &=& \mbox{Pe}\; \tilde{\mathbf{u}}\; \tilde{c}_\pm  -\tilde{\nabla} \tilde{c}_\pm \mp  \tilde{c}_\pm \tilde{\nabla} \tilde{\phi}  \label{eq:bin2}\\
\tilde{c}_- - \tilde{c}_+ &=& 2\tilde{\rho}_s   \label{eq:bin3}
\end{eqnarray}
in terms of the dimensionless variables, $\tilde{x} = x/L$, $\tilde{\nabla}=L\nabla$,  $\tilde{t} = t D/L^2$, $\tilde{c}_\pm = c_\pm/c_0$, $\tilde{\phi} = ze\phi/k_BT$, $\tilde{F}_\pm=F_\pm L/Dc_0$ and $\tilde{\mathbf{u}} = \mathbf{u}/U$, for a geometrical length scale $L$ and characteristic velocity $U$. There are two dimensionless groups that control the solution, the P\'eclet number (ratio of convection to diffusion)
\begin{equation}
\mbox{Pe} = \frac{UL}{D}
\end{equation}
and the dimensionless charge density,
\begin{equation}
\tilde{\rho}_s = \frac{\rho_s}{2zec_0} = \frac{a_p \sigma_s}{2\epsilon_p z e c_0}
\end{equation}
which is the ratio of fixed surface charges to mobile ionic charges per volume, if the pores were filled with a neutral reference solution of salt concentration $c_0$.  As discussed above, the key parameter is $\tilde{\rho}_s$, which determines to what extent the porous medium acts like a ``good membrane" with high conductivity and selectivity for counter-ions ($|\tilde{\rho}_s| \gg 1$). We are mainly interested in ``leaky membranes" with $0< |\tilde{\rho}_s| \ll 1$, which become depleted at high currents, leading to complex nonlinear dynamics.

\begin{figure*}
\begin{center}
(a)
\includegraphics[width=1.9in]{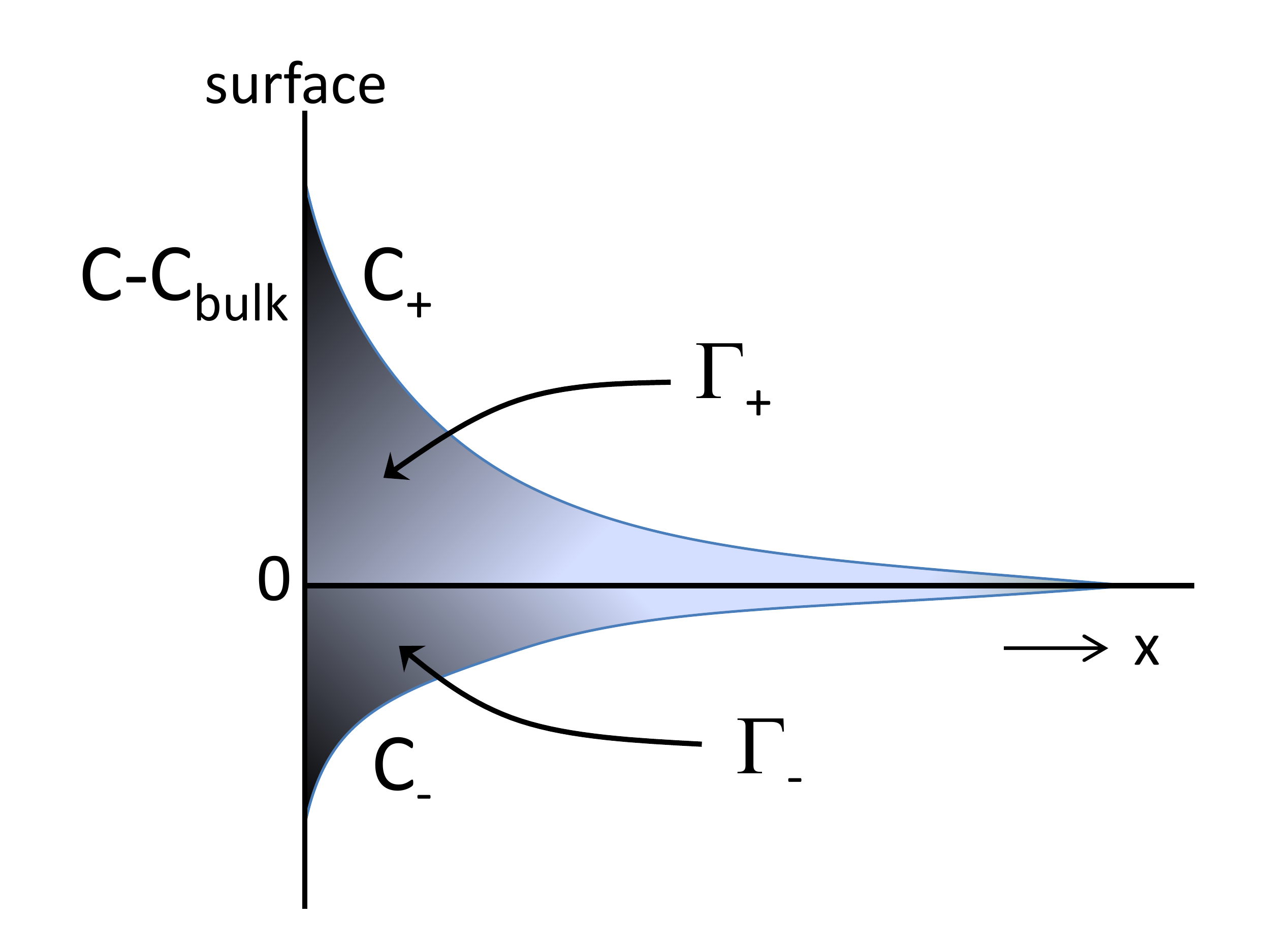}
(b)\includegraphics[width=1.9in]{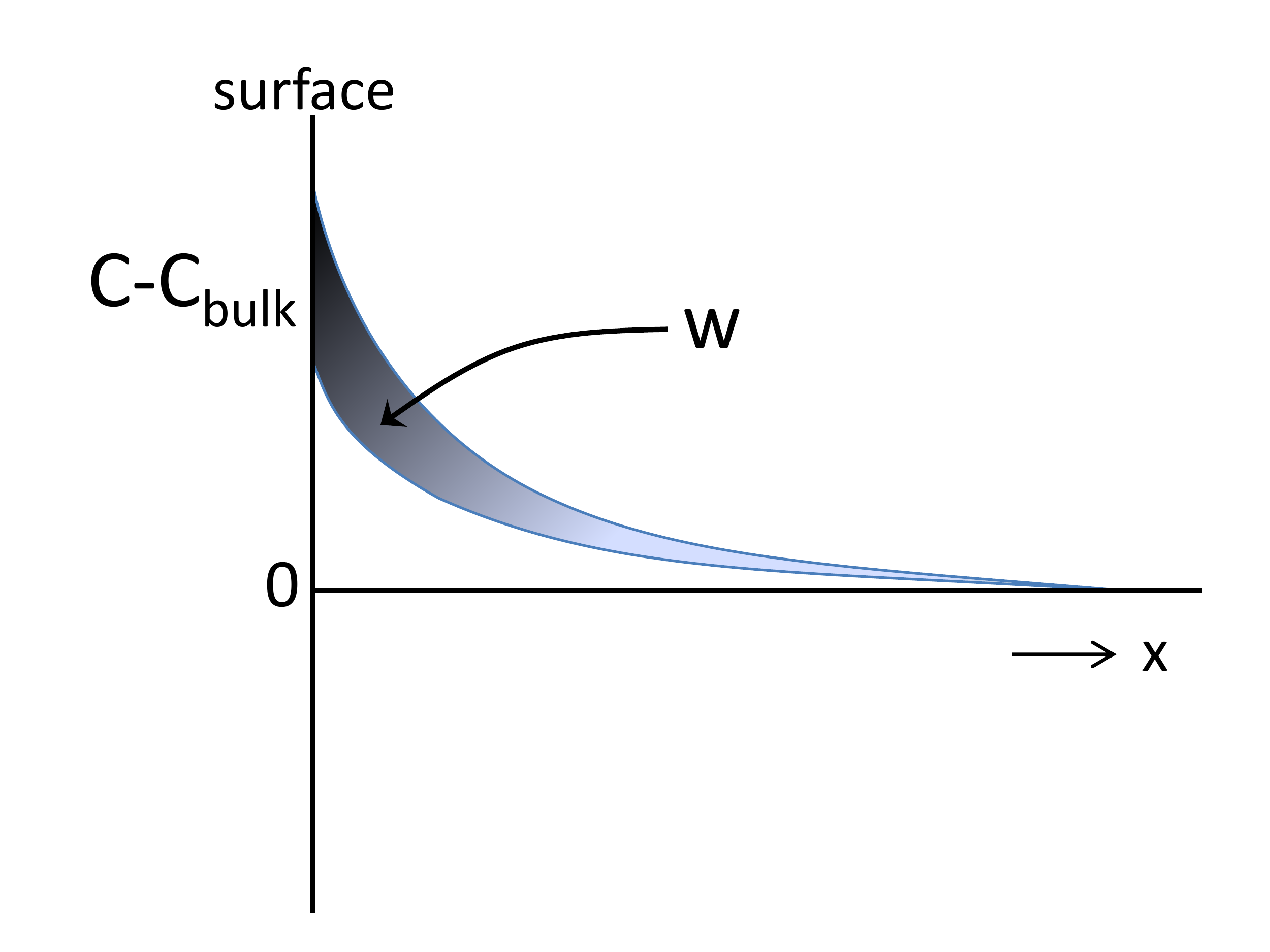}
(c)\includegraphics[width=1.9in]{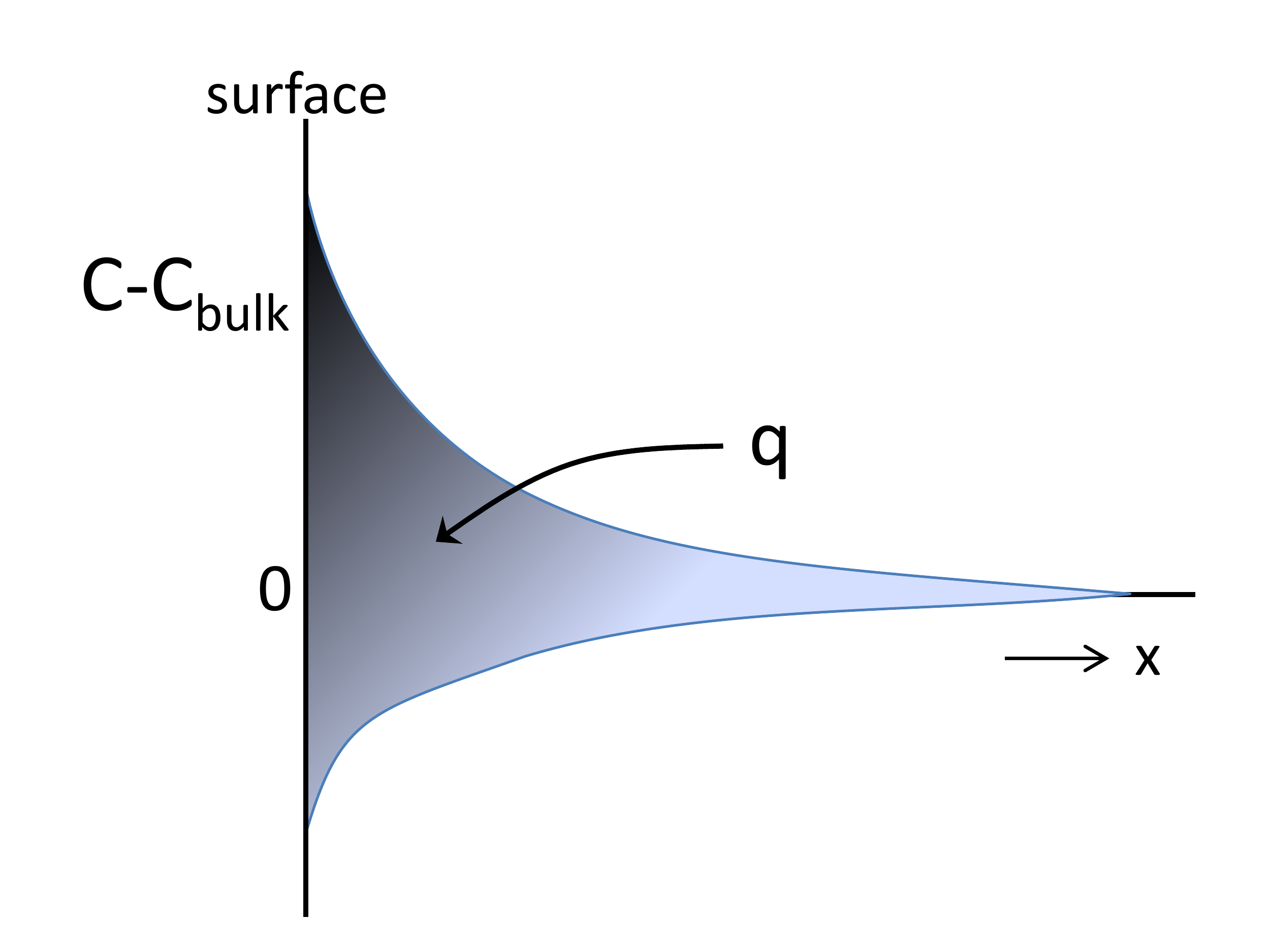}
\caption{Ion Distribution Near a Charged Surface. a) Excess ion concentration, b) Excess neutral concentration, c) Surface charge density}
\label{fig:excess}
\end{center}
\end{figure*}

\subsection*{ {\it Surface Conduction in a Leaky Membrane } }

Before we proceed to the analysis, we comment on the non-standard definition of ``surface conduction" in the LMM~\cite{mani2011desalination}. In classical electrokinetic systems, the term `surface conduction' refers to the excess conduction (from electromigration and electro-osmotic convection) that arises from increased ion concentrations in the electric double layer (EDL) \cite{lyklema1998surface,delgado2007}. Surface conduction in this case can be found by taking the total conduction and subtracting the conduction that would be found in the absence of an EDL. This definition has a long history in electrokinetics from pioneering work of Smoluchowski, Bikerman and Dukhin\cite{smoluchowski1905,bikerman1933ionentheorie,bikerman1935wissenschaftliche,urban1935,bikerman1940,deryagin1969,dukhin1974}. For a particle or pore of characteristic length $a$, the Dukhin number
\begin{equation}
\mbox{Du}=\frac{\kappa_s}{a\kappa_b}
\end{equation}
is the ratio of excess surface conductivity, $\kappa_s$, to bulk conductivity, $\kappa_b$.

While surface conduction is traditionally defined in terms of the excess ion concentration EDL, the more relevant definition for a  leaky membrane is in terms of the total surface charge density, as shown in Fig.~\ref{fig:leaky}. The difference between these two values is shown in Figure~\ref{fig:excess}. Let $\Gamma\pm$ be the total excess surface concentration (Figure~\ref{fig:excess}a). This can be written as 
\begin{equation*}
\Gamma_\pm=\int(c_\pm-c_{bulk})dx
\end{equation*}
for a binary, univalent electrolyte, where $c_-$, $c_+$, and $c_{bulk}$ are the negative, positive, and bulk ion concentrations, respectively, and $x$ is the distance from the charged surface. For a negatively charged surface, $\Gamma_+>0$ and $\Gamma_-<0$. In classical electrokinetics, the excess surface conductivity will come from the excess neutral salt concentration, $w$ \cite{chu2007surface,bazant2004diffuse}.  
\begin{equation*}
w=\int(c_++c_--2c_{bulk})dx=\Gamma_++\Gamma_-
\end{equation*}
In a leaky membrane, however, the total surface ion concentration, or double-layer charge density, $q$, plays an important role and is given by  
\begin{equation*}
q=\int(c_+-c_-)dx=\Gamma_+-\Gamma_- = -\sigma_s
\end{equation*}
Here we consider an EDL at equilibrium and examine a different mechanism for conduction along a charged surface. This surface conduction is in \emph{addition to}, rather than in \emph{excess of}, the conduction through the neutral bulk electrolyte. Throughout this paper, only the surface conduction due to the total surface charge density will be discussed, and will be referred to as SC. 

The importance of surface phenomena generally increases with the surface to volume ratio.  In the case of SC, the role of surface charge is controlled by $\tilde{\rho}_s$,  the ratio of surface charge to bulk ionic charge per volume, which becomes non-negligible in submicron pores, especially at low electrolyte concentrations.  It is important to note that this dimensionless group, while similar, is not the same as the Dukhin number~\cite{mani2011desalination}. The Dukhin number depends on $w$, while $\tilde{\rho}_s$ depends on $\sigma_s=-q$. It is possible for $w$ to be very small while maintaining a large $q$ value, resulting in a large value for $\tilde{\rho}_s$ and a small Dukhin number.

\section*{ Uniform Current without Flow }

\begin{figure}
\begin{center}
\includegraphics[width=3in]{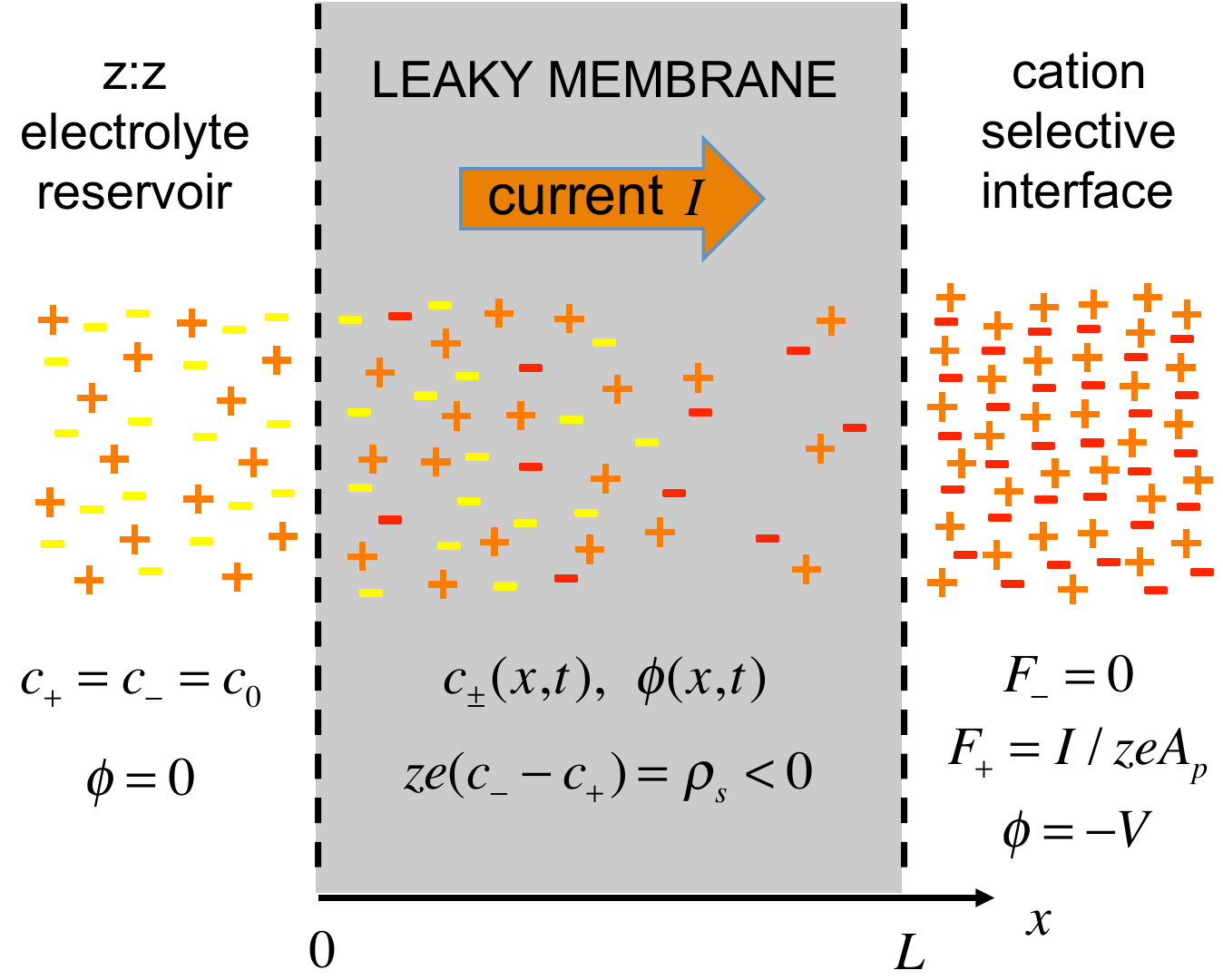}
\caption{ Canonical problem for the Leaky Membrane Model in one dimension.  Symmetric binary electrolyte transport from a reservoir through a weakly cation-selective leaky membrane to an ideally cation-selective surface, such as a (non-leaky) cation-exchange membrane or an electrode undergoing cation electrodeposition.  Solutions appear in Figs. \ref{fig:diff}-\ref{fig:rescaled} for steady and transient applied currents. 
\label{fig:1D} }\end{center}
\end{figure}

In this section, we analyze the canonical problem of ICP in a leaky membrane illustrated in Fig. \ref{fig:1D}.   A symmetric binary electrolyte ($D_\pm=D$, $z_\pm=\pm z$) passes from a reservoir of fixed concentration ($c_\pm=c_0$, $\phi=0$) at $x=0$ through a weakly cation-selective leaky membrane with $\rho_s<0$ through an ideal anion-blocking surface ($F_-=0$, $zeA_p F_+=I$, $\phi=-V$) at $x=L$, which could represent a non-leaky cation-exchange membrane or an electrode consuming cations, e.g. by electrodeposition. We define  $I=A_pJ$, as the total current that passes through the cross-sectional pore area $A_p$, and solve for the transient current-voltage characteristics of the leaky membrane itself, not including interfacial polarization at either end.

\subsection*{ {\it Dimensionless Equations }} 

Following our prior work~\cite{dydek2011overlimiting,mani2011desalination}, it is convenient to transform the LMM for a symmetric, binary electrolyte, Eqs. (\ref{eq:bin1})-(\ref{eq:bin3}), into a dimensionless PDE 
\begin{equation}
\frac{\partial \tilde{c}}{\partial \tilde{t}}=\frac{\partial^2\tilde{c}}{\partial\tilde{x}^2}-\tilde{\rho}_s\frac{\partial^2\tilde{\phi}}{\partial\tilde{x}^2} \label{eq:conctrans}
\end{equation}
for the depletable salt concentration $\tilde{c}=\tilde{c}_-$ (which, as explained above, is equal to the co-ion concentration) and a constraint for the uniform, time-varying current,
\begin{equation}
\tilde{I}=-\left(\tilde{c}-\tilde{\rho}_s\right)\frac{d\tilde{\phi}}{d\tilde{x}} \label{eq:currenttrans}
\end{equation}	
obtained by integrating the cation transport equation.
The dimensionless current 
\begin{equation}
\tilde{I}=\frac{I L}{2zeA_pDc_0},
\end{equation}
is carried by cations and scaled to the limiting current for the case of an ``ideally leaky" membrane, $\tilde{\rho}_s=0$.   The boundary conditions fix the reservoir concentration, $\tilde{c}(0,\tilde{t})=1$, and impose zero anion electro-diffusive flux at the cation-selective surface,
\begin{equation}
\frac{d\ln\tilde{c}}{d\tilde{x}}(1,\tilde{t})=\frac{d\tilde{\phi}}{d\tilde{x}}(1,\tilde{t})   \label{eq:anionbc}
\end{equation}
as well as $\tilde{\phi}(0,\tilde{t})=0$ and $\tilde{\phi}(1,\tilde{t})=-\tilde{V}$, where
\begin{equation}
\tilde{V}=\frac{zeV}{k_BT}   \label{eq:Vt}
\end{equation}
is the dimensionless applied voltage across the leaky membrane (not including interfacial voltage drops and series resistances~\cite{deng2013}).

\subsection*{ {\it Steady State for a Dilute Electrolyte} }

For constant diffusivity in a dilute electrolyte, the steady state can be solved analytically \cite{dydek2011overlimiting}.  The concentration profile is given by an implicit formula,  
\begin{equation}
\tilde{c}-\tilde{\rho}_s\ln(\tilde{c})=1-\tilde{I}\tilde{x}    \label{eq:current}
\end{equation}
and current-voltage relationship
\begin{equation}
\tilde{I}=1-e^{-\tilde{V}}-\tilde{\rho}_s\tilde{V}   \label{eq:IV1D}
\end{equation}
has the form of an equivalent circuit consisting of a diode, representing neutral-electrolyte concentration polarization, in parallel with a shunt resistor, representing surface conduction. 
At low voltages and high bulk conductivity, the model describes the familiar linear Ohmic regime of bulk electrolyte transport,  $\tilde{I}\sim \tilde{V}(1-\tilde{\rho}_s)$. At high voltages and low bulk conductivity, however, the model predicts OLC,   $\tilde{I}\sim 1-\tilde{\rho}_s\tilde{V}$,  with a constant over-limiting conductance sustained by SC.   This formula provides good fit of experimental data for conduction through a silica glass frit (a leaky membrane) from a reservoir to a Nafion membrane in copper sulfate solution~\cite{deng2013}, although the over-limiting conductance also includes strong effects of electro-osmotic flow~\cite{dydek2011overlimiting}.

\subsection*{ {\it Concentration-Dependent Diffusivity } }

In most of our calculations, the diffusivity is treated as a constant, independent of concentration, which is strictly valid only for dilute solutions. In leaky membranes, however, significant concentration variations are possible, which alter the theoretical predictions. It turns out that the steady-state problem can still be solved exactly in implicit form.  Re-writing Eq. (\ref{eq:current}) with a concentration-dependent diffusivity gives
\begin{equation}
\tilde{D}\frac{d\tilde{c}}{d\tilde{x}}-\tilde{D}\tilde{\rho}_s\frac{d\ln(\tilde{c})}{d\tilde{x}}=-\tilde{I},
\end{equation}
with $\tilde{D}=\frac{D(c)}{D(c_0)}$. Integrating by parts and applying the boundary conditions yields
\begin{equation}
\tilde{I}=1-e^{-\tilde{V}}\tilde{D}(e^{-\tilde{V}})-\tilde{\rho}_s\tilde{V}\tilde{D}(e^{-\tilde{V}})+\int^{e^{-\tilde{V}}}_1 (\tilde{c}-\tilde{\rho}_s\ln\tilde{c})\frac{d\tilde{D}}{d\tilde{c}}\,d\tilde{c}.
\label{eq:current_diff}
\end{equation}
If $D$ varies significantly along the channel, the current-voltage relationship will begin to deviate from the ideal case. In particular, overlimiting current will increase if the diffusivity strongly increases with decreasing concentration, which is often the case.

To illustrate effects of non-ideal thermodynamics via Eq.~(\ref{eq:Dconc}), we consider the 
Debye-H\"uckel theory of electrostatic correlations in a  dilute $z:z$ electrolyte. The molar activity coefficient $\gamma$ of each ionic species is given by
\begin{equation}
\ln\gamma=-\frac{(ze)^2}{8\pi\varepsilon kT\lambda_D},
\label{eq:debye}
\end{equation}
where $\varepsilon$ is the permittivity of the solvent and 
\begin{equation}
\lambda_D=\sqrt{\frac{\varepsilon kT}{2(ze)^2c}}
\end{equation}
is the Debye screening length, assumed to be larger than the effective hydrated ion size.  The activity is reduced by attractive interactions between an ion and its screening cloud as the ionic strength is increased, $\ln\gamma \propto - \sqrt{c}$.  The tracer diffusivity $D_i^0$ is taken to be constant, consistent with the moderately dilute range of validity for Debye-Huckel theory.

\begin{figure}[ht]
\centering
\includegraphics[width=\columnwidth,keepaspectratio=true]{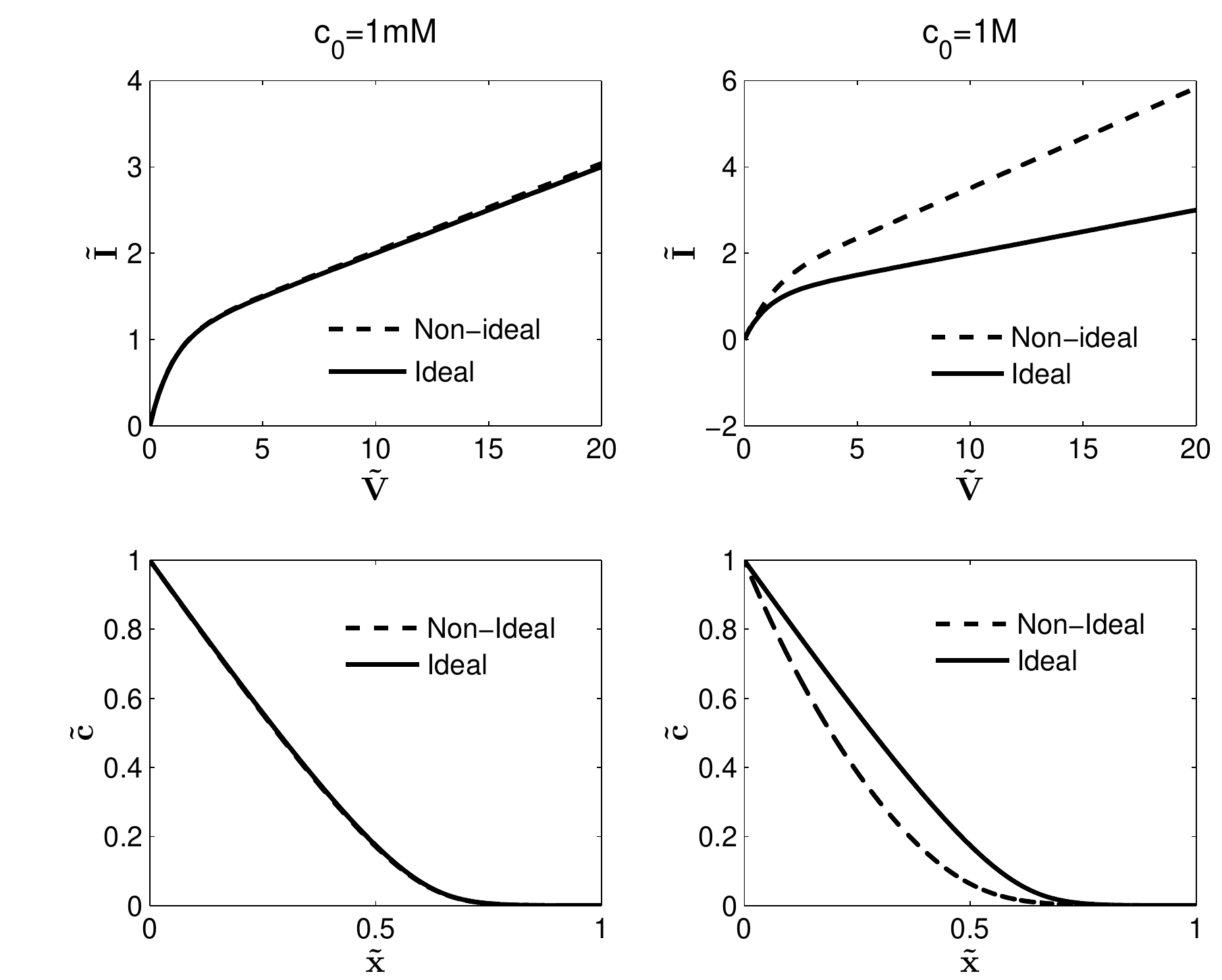}
\caption{ Effects of non-ideal thermodynamics in the model problem of Fig.~\ref{fig:1D}.  Steady-state current-voltage relations  (Top) and concentration profiles (Bottom) are shown for two initial ion concentrations, 1mM (left) and 1M (right), using the concentration-dependent diffusivity $D(c)$ from the  Debye-H\"uckel theory of electrolyte activity.}
\label{fig:diff}
\end{figure}

Using Eqs. (\ref{eq:current_diff}) and (\ref{eq:debye}), the effect of a non-ideal diffusivity (\ref{eq:Dconc}) can be found for varying initial ion concentration. In Figure \ref{fig:diff} we see that for very dilute solutions (1mM) there is very little change in the current-voltage relationship and concentration profile. However, at larger concentrations (1M), there is a significant deviation. The Debye-H\"uckel theory also breaks down and underpredicts the activity, so this example suffices to bound the trends. The overlimiting current is larger than expected from the ideal case and the depletion region is wider. This arises as a result of the diffusivity increasing in the depleted region leading to an increase in mass transfer. This calculation shows that it is generally necessary to account for thermodynamic corrections in the LMM for concentrated electrolytes, although the qualitative results are similar with ideal solution theory, consistent with experimental data~\cite{deng2013}.

\section*{ Mathematics of Deionization Shocks }

\subsection*{ {\it Conductivity Dynamics at Constant Current }}

At constant (over-limiting) current, a leaky membrane (or microchannel~\cite{mani2009propagation}) supports the propagation of deionization shocks~\cite{mani2011desalination}. Below, we will discuss how shocks arise in transient response of a finite system, but we first discuss the mathematics of shock propagation.  It is convenient to recast the equations in terms of the total ion concentration, $\kappa=c_++c_-$, which is proportional to the total conductivity (for equal ionic diffusivities), including both bulk and surface contributions.  Note that this definition of $\kappa$ is different  from the ``bulk" conductivity $\kappa_b$ defined by Mani and Bazant~\cite{mani2011desalination}, set by the depletable co-ion concentration $c=c_-$.

In terms of the dimensionless conductivity $\tilde{\kappa}=\frac{\kappa}{2c_0}$, the LMM takes the form\begin{eqnarray}
\frac{\partial \tilde{\kappa}}{\partial \tilde{t}}=\frac{\partial^2\tilde{\kappa}}{\partial\tilde{x}^2}-\frac{\tilde{\rho}_s\tilde{I}}{\tilde{\kappa}^2}\frac{\partial\tilde{\kappa}}{d\tilde{x}} \label{eq:galv1}  
\end{eqnarray}
with boundary condition
\begin{equation}
 \frac{\partial \tilde{\kappa}}{\partial\tilde{x}}(1,t)+\frac{\tilde{\rho}_s\tilde{I}}{\tilde{\kappa}(1,t)}=-\tilde{I}. \label{eq:galv2}
\end{equation}
The conductivity can be solved independently and the potential then obtained by integrating the current
\begin{equation}
\tilde{I}=-\tilde{\kappa}\frac{\partial \tilde{\phi}}{\partial\tilde{x}}
\end{equation}

By rescaling Eq.~(\ref{eq:galv1}) we can obtain a simple, quasilinear PDE 
\begin{equation}
\frac{\partial\hat{\kappa}}{\partial\tilde{t}}+\frac{\partial(\hat{\kappa}^{-1})}{\partial\tilde{x}}=\frac{\partial^2\hat{\kappa}}{\partial\tilde{x}^2},  \label{eq:burg}
\end{equation}
for the rescaled conductivity $\hat{\kappa}=\tilde{\kappa}/\sqrt{-\tilde{\rho}_s\tilde{I}}$ (assuming that $\tilde{\rho}_s$ is negative). 
This PDE resembles Burgers' equation~\cite{whitham_book,ghosal2010}, where $\hat{\kappa}^2/2$ has been replaced by $\hat{\kappa}^{-1}$. While Burgers' equation can be transformed into the linear diffusion equation by the Cole-Hopf transformation~\cite{cole1951,hopf1950}, there does not seem to be a suitable linearizing transformation for Eq.~(\ref{eq:burg}). However, the scaling used to achieve this simplification suggests that the surface charge density may be as important as the applied current in affecting the resulting conductivity profile. In fact, it will be shown subsequently that $\tilde{\rho}_s$ significantly impacts the conductivity and voltage profiles, bringing about non-linear behavior. 

As with Burgers' equation, the long-time dynamics in an semi-infinite medium is dominated by nonlinear advection (the second term in Eq. ~(\ref{eq:burg})), which leads to shock waves for most boundary and initial conditions. Diffusion plays only a secondary role in determining shock structure~\cite{mani2011desalination}. Neglecting the diffusion term, we obtain a first-order quasilinear PDE,
\begin{equation}
\hat{\kappa}^2\frac{\partial\hat{\kappa}}{\partial\tilde{t}}-\frac{\partial \hat{\kappa}}{\partial\tilde{x}}=0     
\end{equation}
of the same form as Amundson's model of chromatography, Eq.~(\ref{eq:ads}), which can be solved by the Method of Characteristics~\cite{whitham_book,rhee_book}. 

For current in the $+\tilde{x}$ direction, nonlinear kinematic waves in the conductivity profile propagate in the $-\tilde{x}$ direction. For characteristics originating at the end of the leaky membrane where conductivity variations are specified, $\hat{\kappa}(0,\tilde{t}) = \hat{\kappa}_0(\tilde{t})$, the solution for $\tilde{x}<0$ is given by 
\begin{equation}
\hat{\kappa} = \hat{\kappa}_0\left(\tilde{t} + \hat{\kappa}^2 \tilde{x}\right)
\end{equation}
For characteristics originating in the bulk material, the initial conductivity profile, $\hat{\kappa}(\tilde{x},0)=\hat{\kappa}_1(\tilde{x})$, evolves as
\begin{equation}
\hat{\kappa} = \hat{\kappa}_1\left( \tilde{x} + \frac{\tilde{t}}{\hat{\kappa}^2}\right)
\end{equation}
These solutions are valid until characteristics nearly cross, as conductivity gradients become large and lead to shock waves. Multi-valued solutions (``wave breaking") for the conductivity profile are  prevented by diffusion, which stabilizes the shock structure.

\subsection*{ {\it Conductivity Shock Waves }}

A deionization shock is a traveling-wave solution of Eq. (\ref{eq:burg}), $\hat{\kappa}(\tilde{x},\tilde{t}) = f(\xi)$ with $\xi=\tilde{x}-\tilde{v}_s\tilde{t}$, where $\tilde{v}_s$ is the shock velocity.  Let $\hat{\kappa}=f_{-\infty}$ and $\hat{\kappa}=f_{\infty} < f_{-\infty}$ be the conductivity asymptotes ahead ($\tilde{x}\to-\infty$) and behind ($\tilde{x}\to\infty$) the shock, respectively.  The shock profile then satisfies the ordinary differential equation,
\begin{equation}
-\tilde{v}_s f^\prime + (f^{-1})^\prime = f^{\prime\prime}, \ \ f(\pm \infty) = f_{\pm\infty}
\end{equation}
Integrating once we obtain,
\begin{equation}
-\tilde{v}_s (f-f_\infty) + (f^{-1}-f^{-1}_\infty) = f^{\prime}  \label{eq:s2}
\end{equation}
where we impose the boundary condition at $\xi=\infty$, where $f^\prime\to 0$.  If we also impose the boundary condition at $\xi=-\infty$, we obtain the shock velocity (a nonlinear eignevalue):
\begin{equation}
\tilde{v}_{s} = \frac{ f^{-1}_{\infty} -  f^{-1}_{-\infty} }
{ f_{\infty} -  f_{-\infty} } < 0  \label{eq:rh}
\end{equation}
for propagation directed toward the high conductivity region, leaving behind a depleted region behind the shock.  Equation (\ref{eq:rh}) is the Rankine-Hugoniot jump condition expressing integrated mass conservation across the shock.

As in the $\tilde{c}(\tilde{x},\tilde{t})$ formulation~\cite{mani2011desalination}, it is possible to integrate (\ref{eq:s2}) analytically to obtain the shock profile for $\hat{\kappa}(\tilde{x},\tilde{t})$ in implicit form, but it is fairly complicated. A much simpler, approximate solution can be obtained by neglecting the nonlinear advection term ahead of the shock,
\begin{equation}
f(\xi) \approx \left\{ \begin{array}{ll}
f_{-\infty} - (f_{-\infty} - f_\infty ) e^{\xi}  & \mbox{ for }  \xi<0 \\
f_\infty & \mbox{ for }  \xi>0
 \end{array}\right.     \label{eq:sprofile}
 \end{equation}
which corresponds to a truncated exponential profile moving at constant velocity, clearly seen in some of our simulations below (Fig. \ref{fig:conctran}, $\tilde{I}=5$).
This  ``diffusive wave" solution arises whenever an absorbing boundary propagates into a diffusing medium, as in dendritic electrodeposition~\cite{bazant1995}, where metal deposition plays the role of surface conduction in rapidly removing cations from the bulk solution ahead of the wave.   In the absence of flow, deionization shock waves are nonlinearly stable to conductivity perturbations~\cite{mani2011desalination}, due to a mathematical analogy with interface motion in  diffusion-limited dissolution~\cite{bazant2006tld}.

\section*{ Transient Response to a Current Step }

A canonical problem of leaky membrane dynamics is the response to a current step for a dilute, symmetric binary electrolyte in a charge porous material. Three different dynamical regimes in the solution of Eqs. (\ref{eq:conctrans})-(\ref{eq:Vt}) can be identified:

\subsection*{ {\it   Zero Surface Charge: Neutral Electrolytes }}

The limit of zero surface charge, $\tilde{\rho}_s=0$, corresponds to an ``ideally leaky membrane" consisting of quasi-neutral electrolyte confined within the pores. The classical Nernst-Planck equations apply, only with diffusivities corrected for the tortuosity and porosity. An exact series solution can be obtained by finite Fourier transform (as pioneered by Amundson for transport problems):
\begin{equation}
\tilde{c}=1-\tilde{I}\tilde{x}+2\displaystyle\sum_{n=0}^{\infty}\frac{\tilde{I}(-1)^n}{\lambda_n^2} e^{-\lambda_n^2t}\sin{\lambda_nx}, 
\label{eq:fft}
\end{equation}
where $\lambda_n=(n+\frac{1}{2})\pi$. According to Sand~\cite{sand1901}, this classical solution of the diffusion equation was first derived by H. F. Weber in 1879~\cite{weber1879}, who applied it to infer the diffusivity of ZnSO$_4$ from the voltage transient after the interruption of steady current. 
The series can be truncated at a small number of terms without losing accuracy for late times, $\tilde{t} \gg \frac{4}{\pi^2}$. The series is non-uniformly convergent, however, and requires a diverging number of terms at early times. 

For early times or large currents, a more accurate and insightful similarity solution (which effectively sums the series) can be obtained by solving for $\frac{\partial \tilde{c}}{\partial \tilde{x}}$ in a semi-infinite domain. After integrating $\frac{\partial \tilde{c}}{\partial \tilde{x}}$ and applying the boundary conditions, the concentration profile is found to be:
\begin{equation}
\tilde{c}=1+2\tilde{I}\sqrt{\tilde{t}}\left(\eta\operatorname{erfc}\eta-\frac{e^{-\eta^2}}{\sqrt{\pi}}\right), \ \eta=\frac{1-\tilde{x}}{2\sqrt{\tilde{t}}}.
\label{eq:sand}
\end{equation}
This famous result was first obtained by Sand in 1901~\cite{sand1901}, who applied it to infer the diffusivity of CuSO$_4$ from observations of ``Sand's time"~\cite{bard2006electrochemical}, $t_{\text{Sand}}$, the time when the voltage diverges during constant over-limiting current. Solving Eq. (\ref{eq:sand}) for $\tilde{c}(1,\tilde{t})=0$ shows that in this case,
\begin{equation}
\tilde{t}_{\text{Sand}}=\frac{\pi}{4\tilde{I}^2}.
\end{equation}
At Sand's time, the concentration goes to zero at the selective surface, and the potential at that point is undefined and corresponds to an infinite voltage. A crucial observation, however, is that this is voltage spike, a signature of diffusion limitation, can be removed by surface conduction in a leaky membrane.

\subsection*{{\it Large Surface Charge:  Ion Exchange Membranes }}
\begin{figure}
\centering
\includegraphics[width=\columnwidth,keepaspectratio=true]{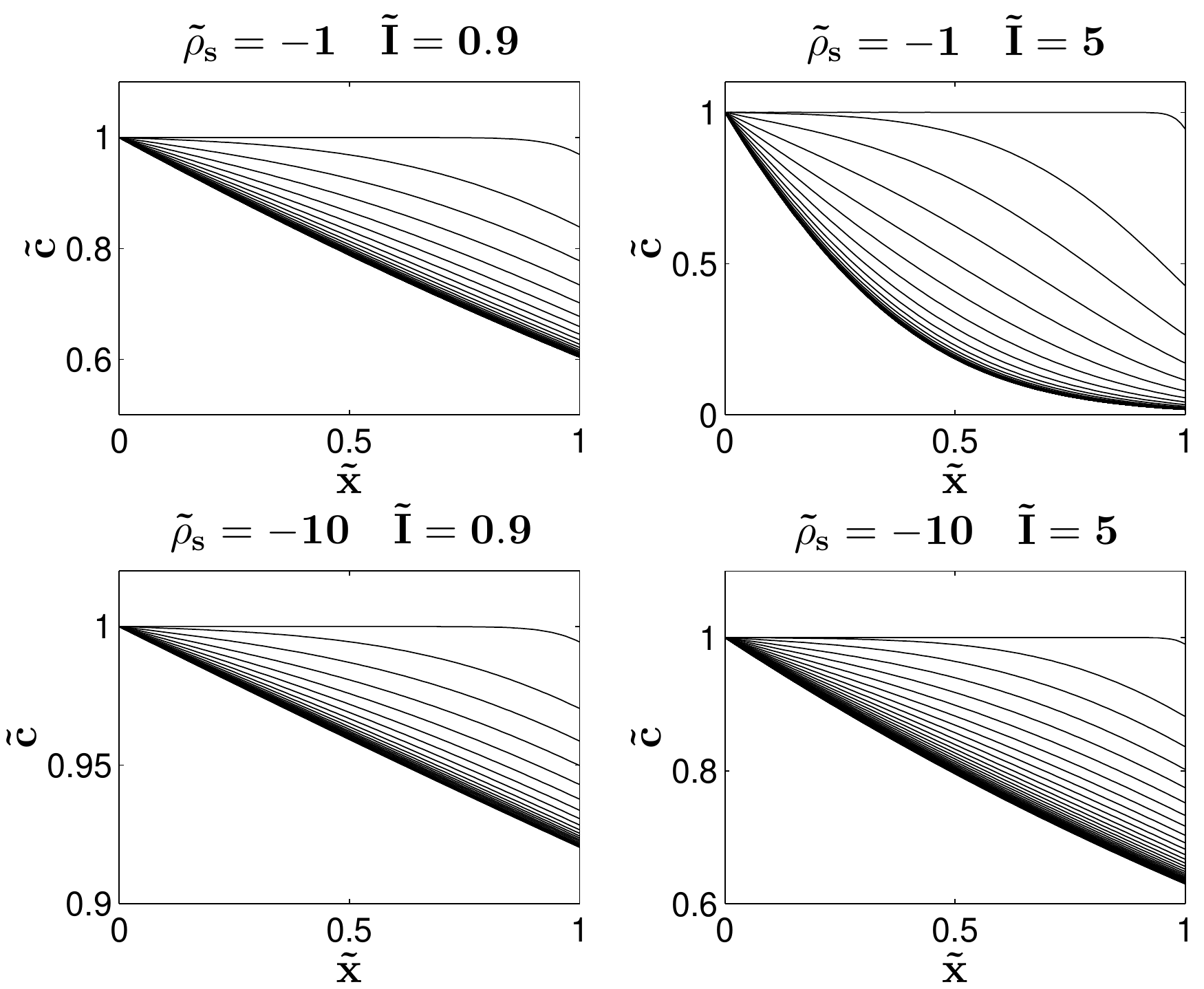}
\caption{Salt concentration evolution after a current step in slightly leaky ion-exchange membranes (Fig. 3) with moderate ($\tilde{\rho}_s=-1$) and high ($\tilde{\rho}_s=-10$) negative surface charge, just below ($\tilde{I}=0.9$) and far above ($\tilde{I}=5$) the limiting current. }
\label{fig:highrho}
\end{figure}

The other extreme is the case of $\tilde{\rho}_s\gg1$, which corresponds to a highly charged ion-exchange membrane with high electrochemical permselectivity. As shown in Figure~\ref{fig:highrho}, high values of surface charge suppress large concentration gradients, even under OLC. This lack of concentration gradient results in a nearly constant potential across the system, meaning that under $\tilde{\rho}_s\gg1$ conditions the system is almost a purely controlled by the diffusion of the counter ions. Interestingly, the transient behavior of the case of very large $\tilde{\rho}_s$ behaves similarly to that of the case of $\tilde{\rho}_s=0$ (as shown in the FFT solution), where transport is dominated by diffusion in the absence of a significant potential gradient.  In a neutral medium ($\tilde{\rho}_s=0$)  the ambipolar diffusivity (based on the diffusivities of the counter and co-ions) determines the transient behavior, but the counter-ion diffusivity alone dominates at high charge density ($\tilde{\rho}_s\gg1$).

\subsection*{{\it Small Surface Charge: Leaky Membranes}}
\begin{figure}
\centering
(a)\includegraphics[width=2.9in,keepaspectratio=true]{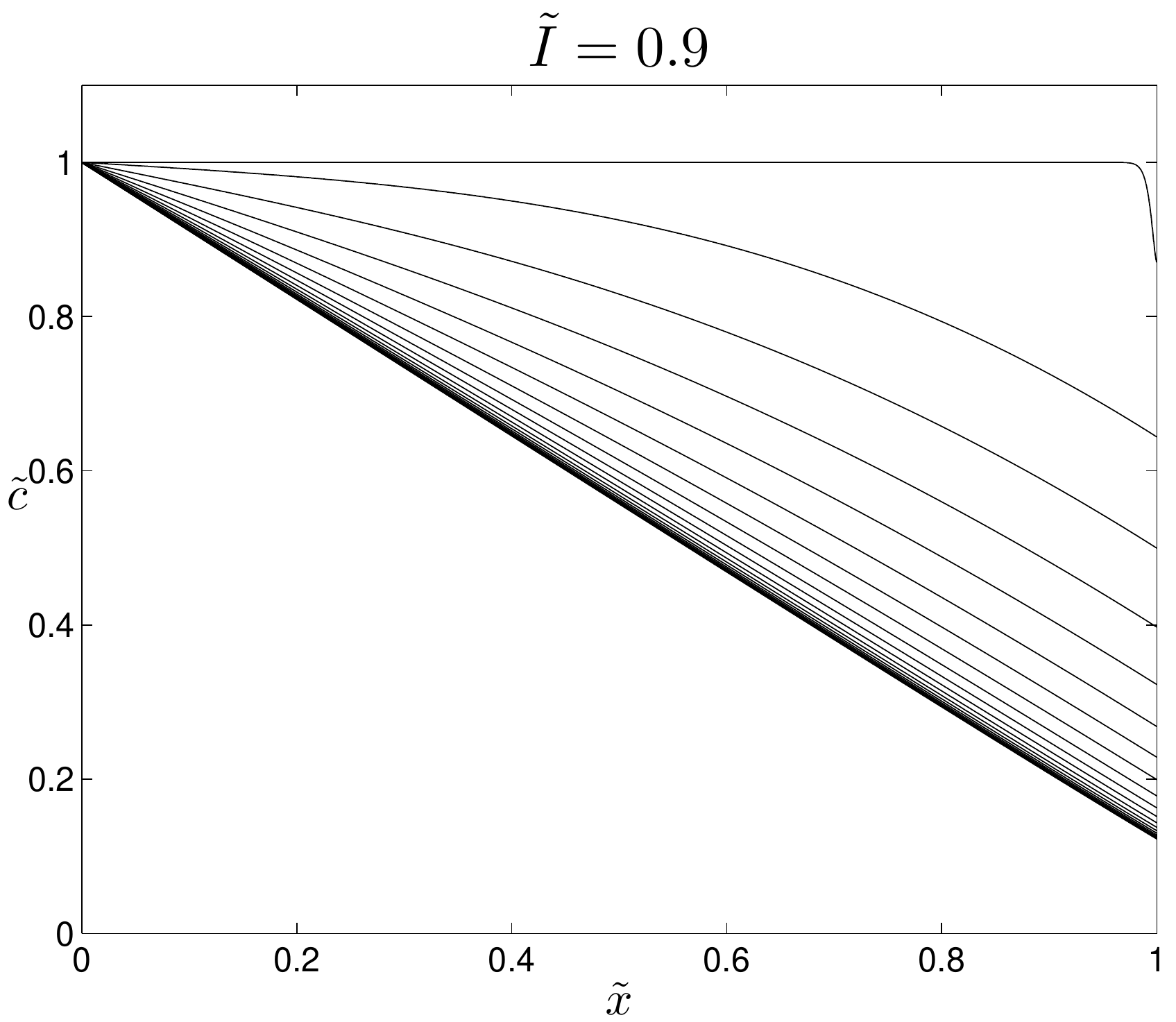}\\
(b)\includegraphics[width=2.9in,keepaspectratio=true]{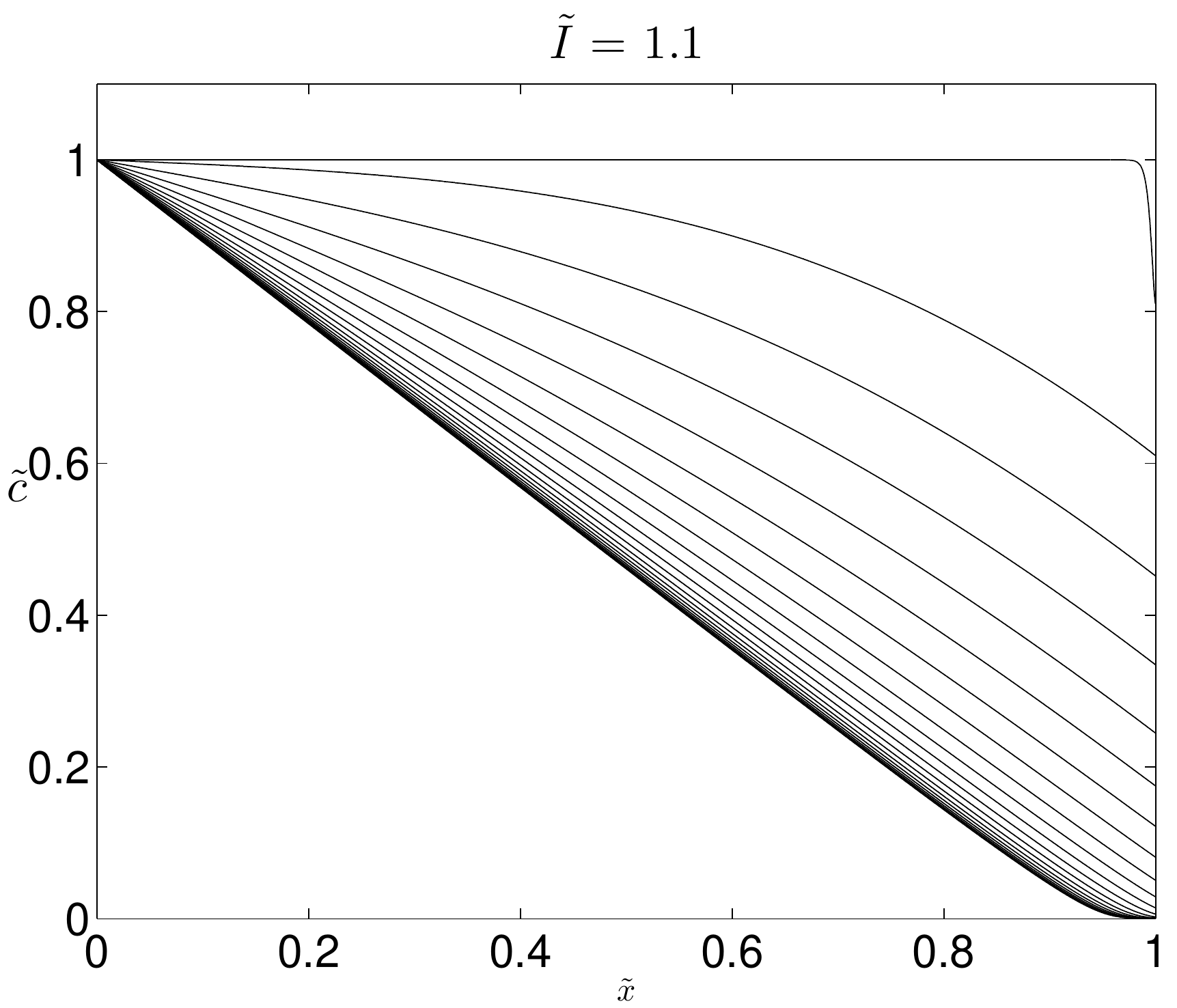}\\
(c)\includegraphics[width=2.9in,keepaspectratio=true]{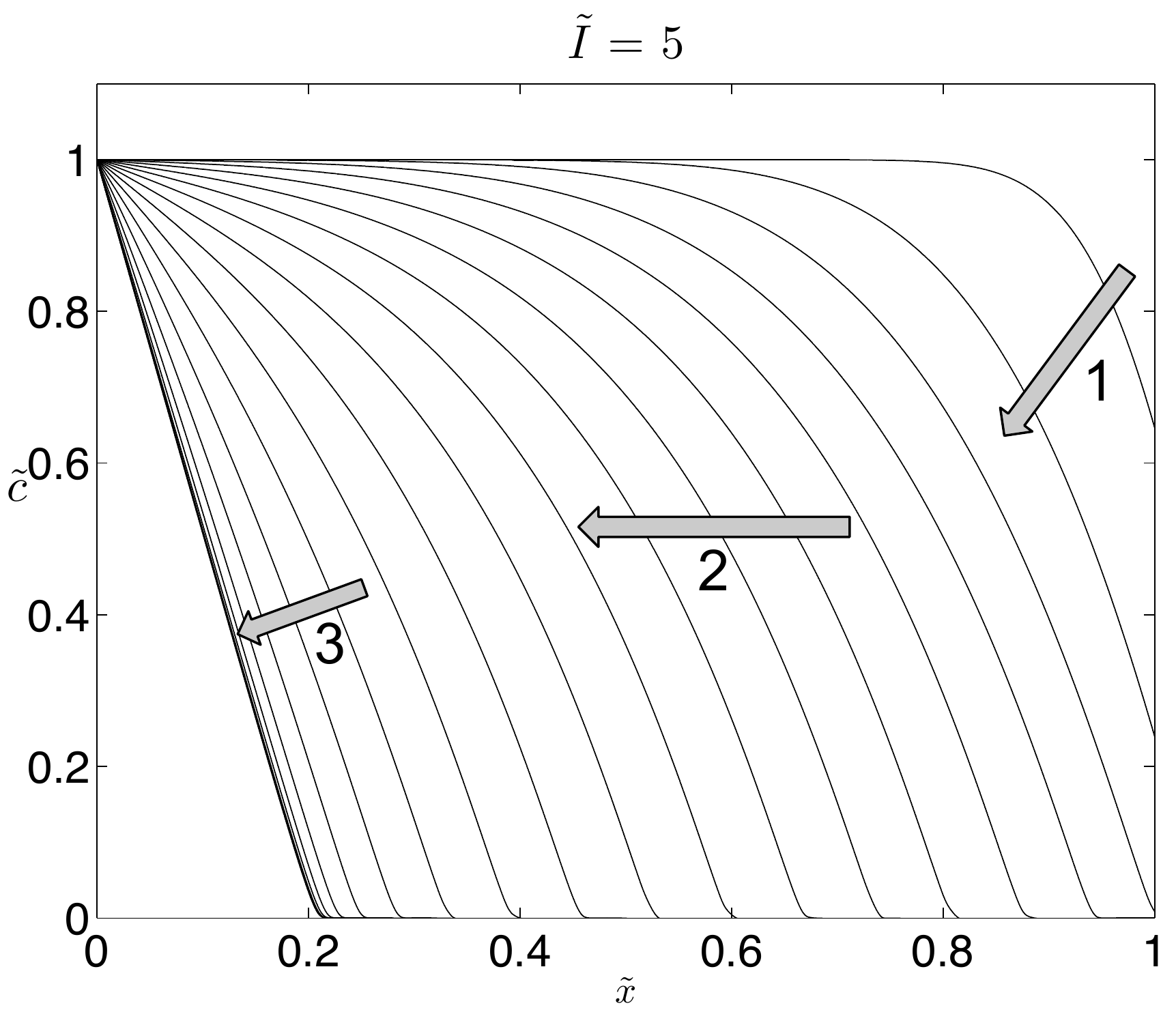}
\caption{ Transient response of the salt concentration  to a current step across a leaky membrane ($\tilde{\rho}_s=-0.01$, Fig. 3). Just below the limiting current (a), the dynamics is dominated by linear diffusive relaxation, while nonlinearity begins to alter the concentration profile just above the limiting current (b). Well above the limiting current (c), the initial diffusion layer drives total salt depletion (stage 1), followed by a new dynamical regime of shock propagation (stage 2), ending in relaxation to steady state (stage 3).
}
\label{fig:conctran}
\end{figure}

In a leaky membrane, the surface charge is relatively small, $\tilde{\rho}_s = O(1)$ but plays an important role. Below the limiting current ($\tilde{I}\ll 1$), a small dimensionless surface charge ($0\leq |\tilde{\rho}_s|\ll 1$) acts as a regular perturbation of the system, and the solution to our model problem remains close to the diffusive relaxation of a neutral electrolyte (\ref{eq:fft}) for all times. Above the limiting current ($\tilde{I}>1$), however, even a very small,  but non-zero, surface charge ($0<|\tilde{\rho}_s|\ll 1$) acts as a singular perturbation that significantly alters the dynamics. The transient concentration profile in our model problem for three different currents is shown in Figure~\ref{fig:conctran}, and several OLC voltage responses are given in Figure \ref{fig:volttrans}. Under OLC conditions, the ion concentration profile undergoes three stages: 1) Depletion, 2) Shock Propagation and 3) Relaxation.

\begin{enumerate}
\item {\bf Salt Depletion. }
During this early stage the ion concentration is depleted at the selective surface and behaves similarly to the classic model ($\tilde{\rho}_s=0$). This is more clearly shown in second half of Figure \ref{fig:volttrans} where time has been rescaled with respect to the classically derived Sand time, $t_{\text{Sand}}$. With this rescaling it is clear that the large voltage increase, corresponding to full depletion, occurs at $\tilde{\tau}_1 = \tilde{t}_{\text{Sand}}$. The fact that time scales for the classical case still apply when SC is taken into consideration is further shown by demonstrating the impact of $\tilde{\rho}_s$ in Figure~\ref{fig:volttransrho}. In this figure the voltage response is shown for decreasing surface charge. As the absolute value of $\tilde{\rho}_s$ decreases an order of magnitude the voltage drop increases about an order of magnitude. As the dimensionless surface charge continues to decrease the system grows closer to the classical result, as expected.

\begin{figure}
\centering
(a)\includegraphics[width=2.8in,keepaspectratio=true]{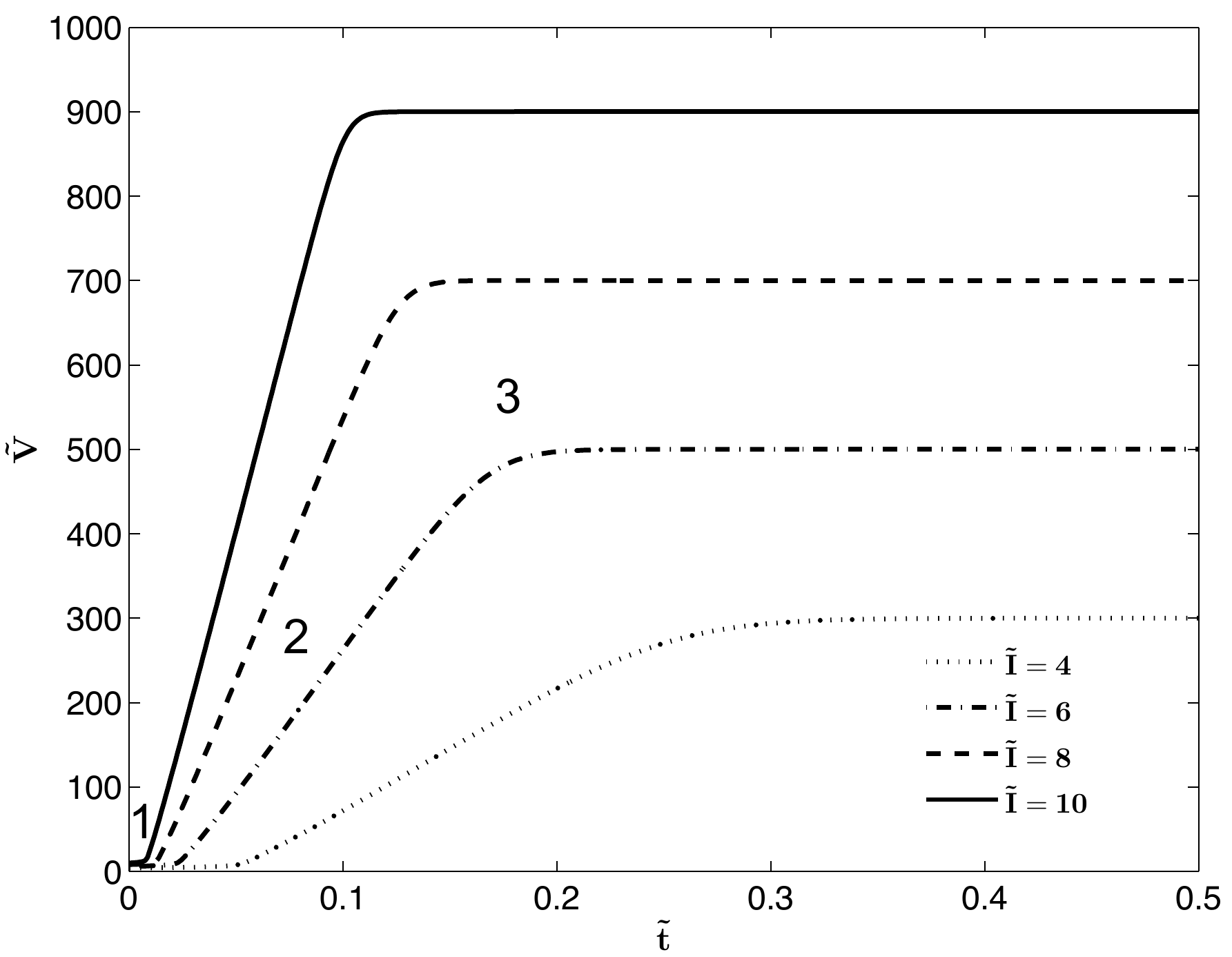}\\
(b)\includegraphics[width=2.8in,keepaspectratio=true]{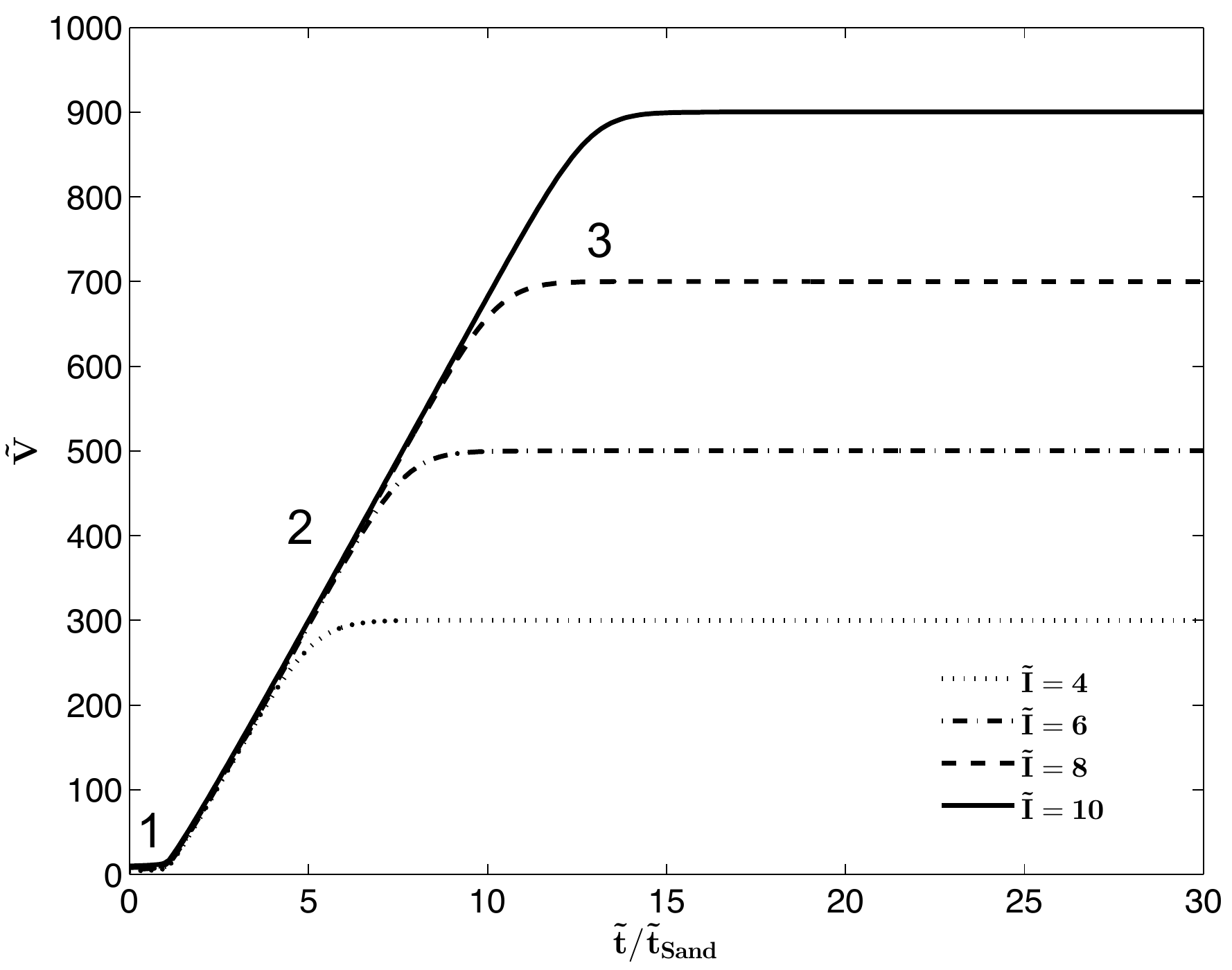}
\caption{Transient voltage of across a leaky membrane in response to a current step with increasing current ($\tilde{\rho}_s=-0.01$). Top: Dimensionless voltage versus dimensionless time.  Bottom: After rescaling time to Sand's time, three distinct dynamical regimes (Fig. \ref{fig:conctran}) appear at high current: 1. depletion, 2. shock propagation, and 3. relaxation to steady state.  }
\label{fig:volttrans}
\end{figure}

\begin{figure}
\centering
\includegraphics[width=3in,keepaspectratio=true]{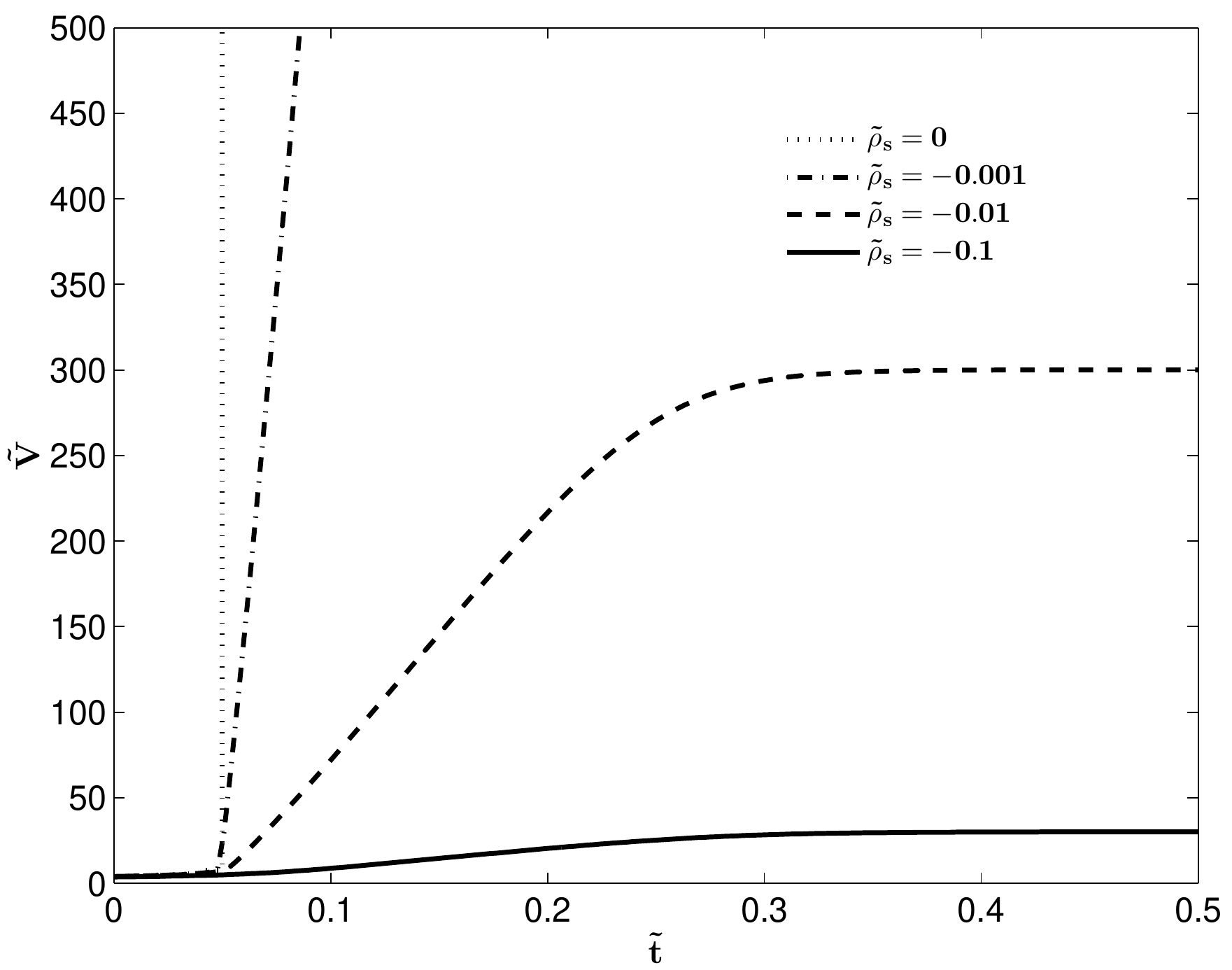}
\caption{The effect of varying the surface charge on the transient voltage response to a current step (Fig. \ref{fig:volttrans}) for $\tilde{I}=4$.}
\label{fig:volttransrho}
\end{figure}

\item {\bf Shock Propagation}. After the co-ion concentration is fully depleted at $\tilde{x}=1$, a deionization shock appears and propagates away from the selective surface~\cite{mani2009propagation} with a stable exponential profile~\cite{mani2011desalination}, given by Eq. (\ref{eq:sprofile}). In the case of total salt depletion behind the shock ($\tilde{\kappa}_\infty=-\tilde{\rho}_s\ll 1$) and unperturbed conductivity ahead ($\tilde{\kappa}_{-\infty}=1-\tilde{\rho}_s>1$), the shock velocity, Eq. (\ref{eq:rh}), takes the simple form
\begin{equation}
\tilde{v}_s = - \frac{\tilde{I}}{1-\tilde{\rho}_s}
\end{equation}
which is equal to the (dimensionless) electromigration velocity of co-ions~\cite{bazant1995,mani2011desalination}, as required by mass conservation across the shock, since no co-ions are left behind. 

Over the duration of stage 2 $\tau_2$ can be estimated as the time for the shock to move at velocity $\tilde{v}_s$ from the selective surface at $\tilde{x}=1$ to the edge of the steady-state depletion region at $\tilde{x}=\tilde{I}^{-1}$. This implies the scaling:
\begin{equation}
\tilde{\tau}_2 = \left( \tilde{I}^{-1} - \tilde{I}^{-2} \right)(1 - \tilde{\rho}_s)
\end{equation}

Next we analyze the transient voltage during stage 2. 
The dimensionless conductivity in the depleted region is approximately $\tilde{\rho}_s\ll 1$, which dominates the total electrical resistance of the leaky membrane.  The length of the depleted region at any time past $t_{\text{Sand}}$ is equal to the shock velocity ($\tilde{I}$) times time. Therefore the resistance of the depleted region is $\tilde{I}(\tilde{t}-\tilde{t}_{Sand})/\tilde{\rho}_s$. Using Ohm's law, the voltage thus scales as  
\begin{equation}
\tilde{V} \sim \frac{\tilde{I}^2(\tilde{t}-\tilde{t}_{Sand})}{\tilde{\rho}_s} . 
\end{equation}
This scaling is verified at high currents in Figure \ref{fig:rescaled}, where $\tilde{V}\tilde{\rho}_s/\tilde{I}$ is plotted against $\tilde{I}(\tilde{t}-\tilde{t}_{Sand})$, leading to a data collapse of both stages 2 and 3 of the dynamics. As the applied current increases, thereby strengthening the shock, the system closely obeys these scaling laws. At larger currents the depleted region nearly encompasses the entire system length with the shock propagating for a interval scaling as $\tilde{\tau}_2 \sim \tilde{I}^{-1}$ after Sand time has been achieved. 

\begin{figure}
\centering
\includegraphics[width=2.8in,keepaspectratio=true]{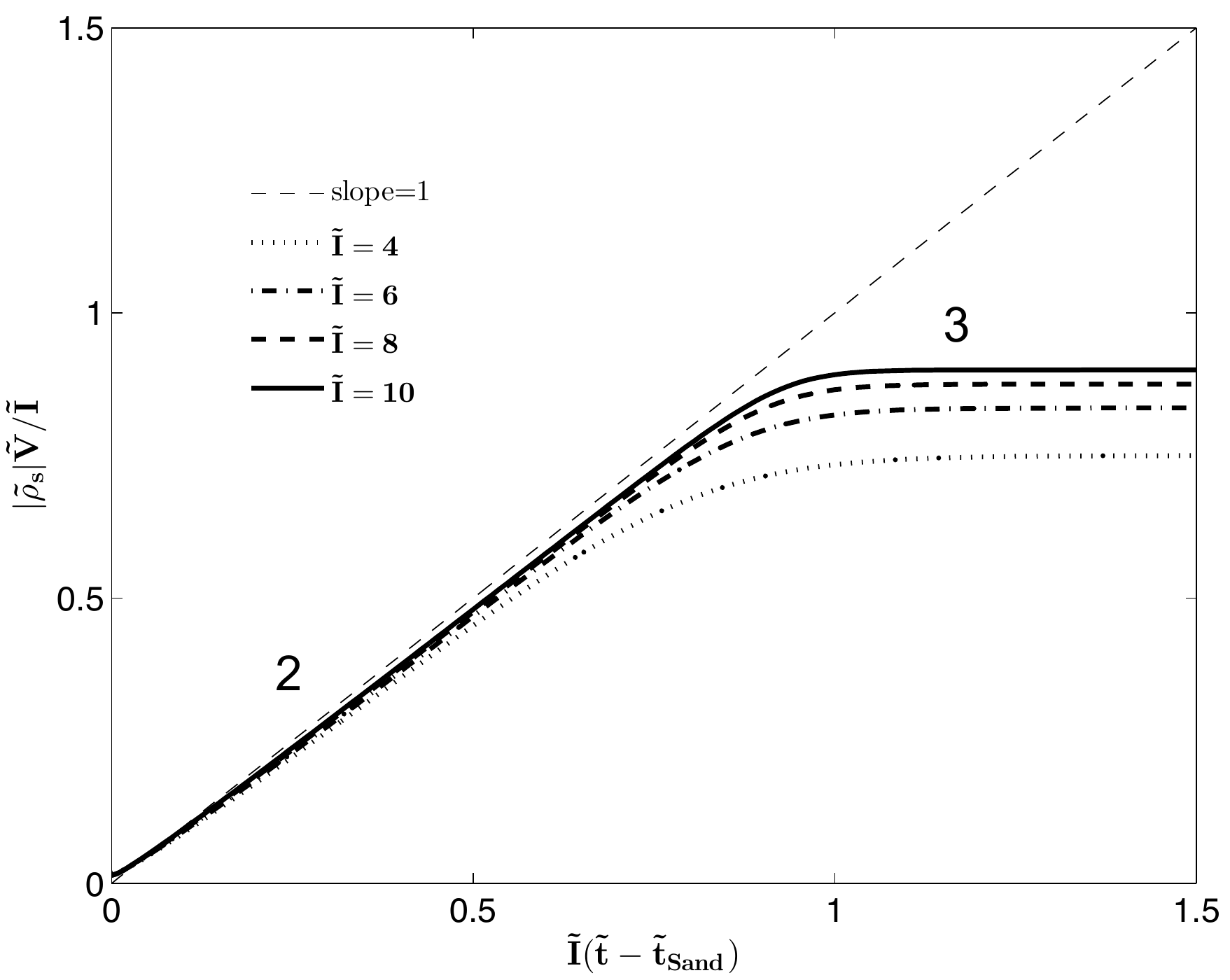}
\caption{Scaling and data collapse of numerical solutions for the transient voltage (Fig. \ref{fig:volttrans}) across stage 2 (shock propagation) and stage 3 (relaxation to steady state) for different applied currents. ($\tilde{\rho}_s=-0.01$)}
\label{fig:rescaled}
\end{figure}

{\it Aside: Shock Propagation at Constant Voltage. } While the depletion region grows linearly with time under constant current conditions, it has been shown that depletion propagates as $t^{1/2}$ under constant-voltage conditions \cite{zangle2010constantvoltage}. This scaling can be easily revealed by estimating the shock as a moving step function, with a depletion region length equal to $\delta(\tilde{t})$. From Eq. (\ref{eq:galv2}) and the boundary conditions the voltage drop across the system is given as $\tilde{V}=\tilde{I}(t)\int_0^1\frac{1}{\tilde{\kappa}}d\tilde{x}$. Integrating over the step function gives $\tilde{V}\approx\tilde{I}[(1-\delta)/\tilde{\kappa}_{-\infty}+\delta/\tilde{\kappa}_\infty]$. Since $\tilde{\kappa}_\infty\ll\tilde{\kappa}_{-\infty}$, $\tilde{V}\approx-\tilde{I}\delta/\tilde{\rho}_s$. The size of the depletion region is equal to the integral of the shock velocity, or $\tilde{I}=d\delta/d\tilde{t}$. Combining this with the voltage drop approximation, the depletion region length is found to propagate as 
\begin{equation}
\delta(\tilde{t})\approx\sqrt{-2\tilde{\rho}_s\tilde{V}\tilde{t}} \ \   \mbox{ (constant voltage) }
\end{equation}
demonstrating the $t^{1/2}$ behavior.

\item {\bf Relaxation to Steady State.} Under moderate surface charge, SC only plays a dominant role when ion concentrations are very low, such as in the depletion region. Outside of this region, transport is dominated by linear diffusion. Once the shock is close to its final position (determined by the applied current or voltage) the concentration profile  relaxes to the steady-state profile. As diffusion is the dominant transport mechanism, the transient scaling during this stage will be similar to the $\tilde{\rho}_s=0$ case, solved earlier by FFT.  The main difference is that the relevant length scale is not the total leaky membrane thickness, but rather the width of the steady-state diffusion layer $\tilde{I}^{-1}$. As such the eigenvalues are rescaled to $\bar{\lambda}_n \approx \left(n+\frac{1}{2}\right)\pi \tilde{I}$, and thus the time scale for relaxation is 
\begin{equation}
\tilde{\tau}_3 = \frac{1}{\bar{\lambda}_1^2} = \frac{4}{\pi^2 \tilde{I}^2}
\end{equation}
Finally, we are able to predict the total time to steady state by adding the times for all three stages
\begin{equation}
\tilde{\tau} %= \tilde{\tau}_1 + \tilde{\tau}_2 +\tilde{\tau}_3 
= (1-  \tilde{\rho}_s)\tilde{I}^{-1} + \left( \frac{\pi}{4} + \frac{4}{\pi^2} - 1 + \tilde{\rho}_s \right) \tilde{I}^{-2}
\end{equation}
In the limit of large currents, $\tilde{I} \gg 1$, the response time is dominated by the time for the shock to cross the full thickness of the leaky membrane (first term).

\end{enumerate}

\section*{ Steady State with Normal Flow }

\begin{figure}
\centering
\includegraphics[width=3in,keepaspectratio=true]{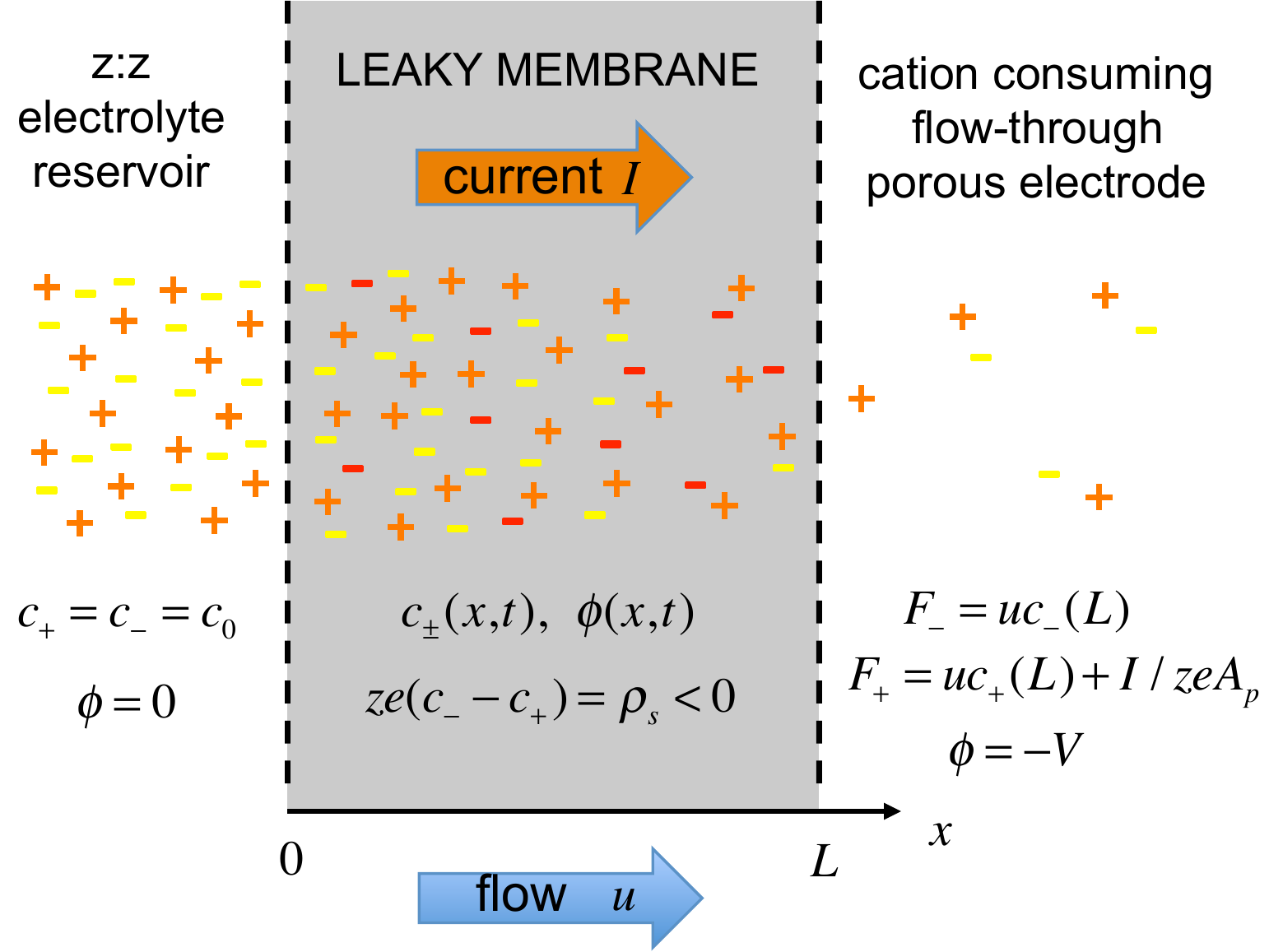}
\caption{Sketch of the model problem flow aligned with the current, normal to the cation selective surface. Current is passed from a reservoir through a leaky membrane to a flow-through porous electrode that consumes cations, but allows neutral salt to pass by convection.}
\label{fig:1dflow}
\end{figure}

In the previous section we examined how OLC creates ICP by forcing ion depletion regions to develop. In this section we explore the effects of a uniform normal flow $u=U$ directed toward the selective surface on the depletion region and compute steady-state concentration profiles and current-voltage characteristics. This idealized situation shown in Fig.\ref{fig:1dflow} is relevant for flow-through porous electrodes~\cite{suss2012capacitive}, as well as dominant normal flow that exits through a small side outlet near the membrane~\cite{deng2013}.  Instead of a solid right wall, a perfect porous electrode is in place at $x=L$, which allows fluid and neutral salt to pass through while removing all excess cations. 

%\begin{equation}
%\begin{split}
%u\frac{dc_+}{dx}=D\left[\frac{d^2c_+}{dx^2}+\frac{ze}{k_BT}\frac{d}{dx}\left(c_+\frac{d\phi}{dx}\right)\right],\\ u\frac{dc_-}{dx}=D\left[\frac{d^2c_-}{dx^2}-\frac{ze}{k_BT}\frac{d}{dx}\left(c_-\frac{d\phi}{dx}\right)\right],
%\label{eq:normal flow}
%\end{split}
%\end{equation}
%\begin{equation}
%\begin{split}
%I_{\text{electrode}}=I_++&I_-, \
%\frac{I_+}{ze}=-D\left(\frac{dc_+}{dx}+\frac{ze}{k_BT}c_+\frac{d\phi}{dx}\right) ,\\ &\frac{I_-}{ze}=D\left(\frac{dc_-}{dx}-\frac{ze}{k_BT}c_-\frac{d\phi}{dx}\right)=0.
%\end{split}
%\label{eq:normal i}
%\end{equation}

\subsection*{ {\it Concentration Profiles and Polarization Curves }}

Starting from Eqs.  (\ref{eq:bin1})-(\ref{eq:bin3}), a pair of dimensionless equations for the steady state is achieved by averaging over the cross section,
\begin{eqnarray} 
\mbox{Pe}\frac{d\tilde{c}}{d\tilde{x}}&=&\frac{d^2\tilde{c}}{d\tilde{x}^2}-\tilde{\rho}_s\frac{d^2\tilde{\phi}}{d\tilde{x}^2} \label{eq:normal1} \\
0 &=& \frac{d}{d\tilde{x}}\left[\left(\tilde{c}-\tilde{\rho}_s\right)\frac{d\tilde{\phi}}{d\tilde{x}}\right]    
\end{eqnarray}
 where the P\'eclet number $\mbox{Pe}=UL/D$ controls the importance of convection relative to diffusion. Note that convection drops out of the charge balance (second equation) because we assume a constant  surface charge density.  For a flow-though electrode, the boundary conditions are more subtle. Only neutral salt can pass through the electrode, and therefore, to the right of $x=L$, $uc_+$ must equal $uc_-$. However, charge conservation within the membrane forces $c_+>c_-$ between $x=0$ and $x=L$. As a result, a streaming current develops, $I_{stream}=-U\rho_s$, that accounts for this imbalance. The total current is the sum of the electro-diffusive fluxes, discussed above, and the streaming current: 
\begin{equation}
\tilde{I}=-\left(\tilde{c}-\tilde{\rho}_s\right)\frac{d\tilde{\phi}}{d\tilde{x}}-\mbox{Pe}\tilde{\rho}_s
\label{eq:normal2}
\end{equation}
The anion electro-diffusive flux also vanishes, Eq. (\ref{eq:anionbc}).

The equations can also be rearranged to determine the dimensionless total ion concentration (or conductivity), $\tilde{\kappa}=\tilde{c}-\tilde{\rho}_s$,  instead of $\tilde{c}$:\begin{eqnarray}
\mbox{Pe}\frac{d\tilde{\kappa}}{d\tilde{x}}&=&\frac{d^2\tilde{\kappa}}{d\tilde{x}^2}-\frac{\tilde{\rho}_s(\tilde{I}+\mbox{Pe}\tilde{\rho}_s)}{\tilde{\kappa}^2}\frac{d\tilde{\kappa}}{d\tilde{x}}\\
%0 &=& \frac{d}{d\tilde{x}}\left[\tilde{\kappa}\frac{d\tilde{\phi}}{d\tilde{x}}\right] \\
-(\tilde{I}+\mbox{Pe}\tilde{\rho}_s)&=&\frac{d\tilde{\kappa}}{d\tilde{x}}(1) +\frac{\tilde{\rho}_s(\tilde{I}+\mbox{Pe}\tilde{\rho}_s)}{\tilde{\kappa}(1)}  \\
\tilde{\kappa}(0) &=& 1-\tilde{\rho}_s
%\frac{d\tilde{\kappa}}{d\tilde{x}}(1) &=& \left(\tilde{\kappa}(1) + \tilde{\rho}_s\right) \frac{d\tilde{\phi}}{d\tilde{x}}(1)
\end{eqnarray}	
Once this boundary value problem is solved for $\kappa(x)$, the potential profile and voltage are obtained from the current relation by a simple integration:
\begin{equation}
\tilde{\phi}(x) = -(\tilde{I}+\mbox{Pe}\tilde{\rho}_s)
\int_0^{\tilde{x}}  \frac{ ds }{ \tilde{\kappa}(s) }
\end{equation}
With flow, the analytical solution becomes more challenging, so concentration profiles and current-voltage relationships are found numerically.

\begin{figure}
\centering
(a)\includegraphics[width=3in]{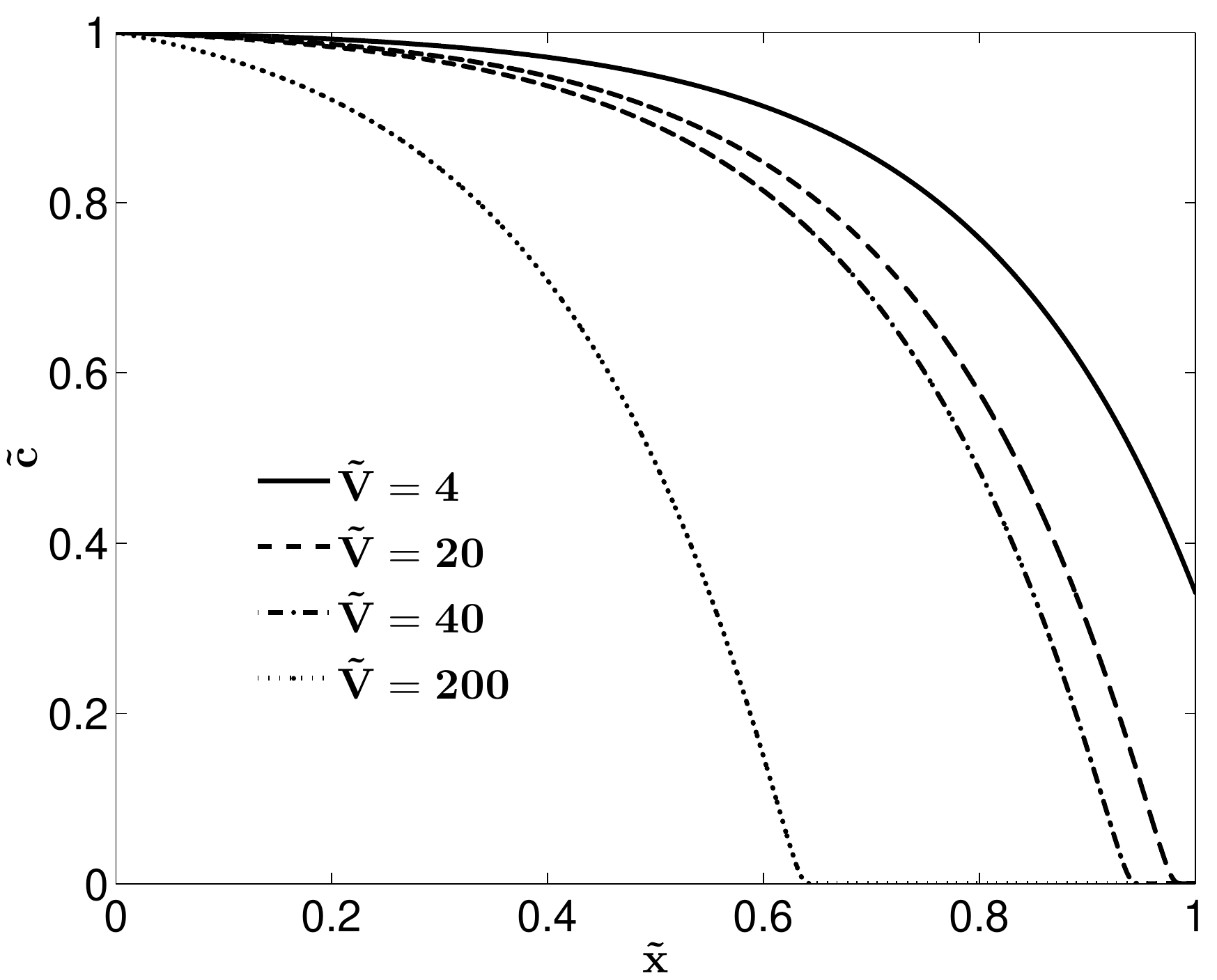}\\
(b)\includegraphics[width=3in]{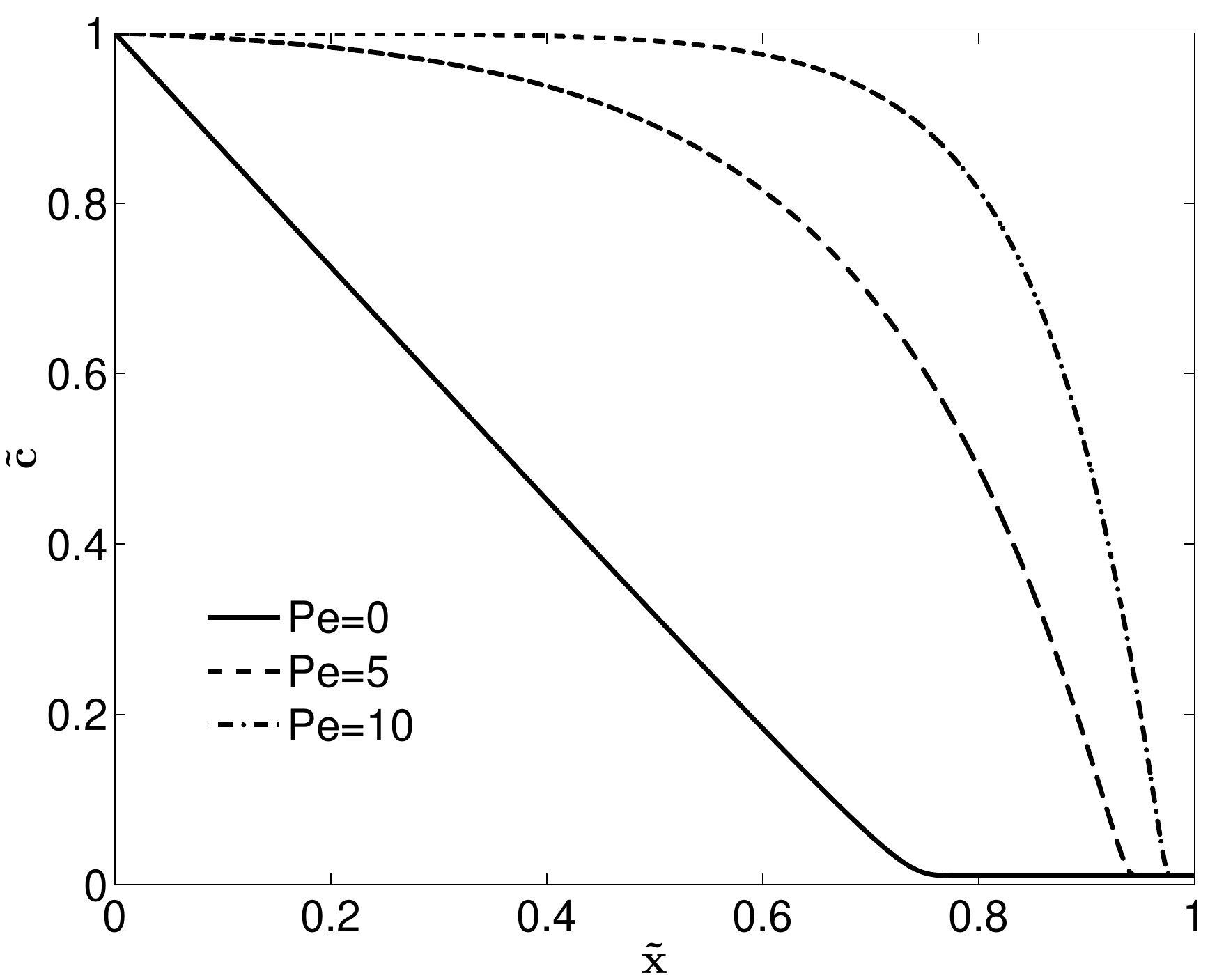}
\caption{Concentration profiles in a leaky membrane ($\tilde{\rho}_s=-0.01$) with normal flow (Fig.~\ref{fig:1dflow}) while varying (a) the voltage with a high flow rate,  $\mbox{Pe}=5$, or (b) the flow rate at high voltage, $\tilde{V}=40$. }
\label{fig:1Dflowconc}
\end{figure}

In Figure~\ref{fig:1Dflowconc} several concentration profiles are shown for varying voltage and $\mbox{Pe}$ values. As shown previously\cite{dydek2011overlimiting}, increasing the applied voltage increases the amount of depletion. The addition of convection pushes all ions toward the membrane, and the linear concentration profile of the quasi-steady diffusion layer gives way to the exponential  profile of a diffusive wave, propagating against the flow~\cite{bazant1995,mani2011desalination}. As the flow rate increases the concentration profile appears more shock-like with a decreasing shock width. Additionally, as the flow rate increases, the depletion region shrinks, requiring a higher applied voltage to maintain its size. At high $\mbox{Pe}$ the steady-state concentration profile converges  to the propagating shock solution, Eq. (\ref{eq:sprofile}), where the uniform flow holds the shock in place.

\begin{figure}
\centering
\includegraphics[width=3in]{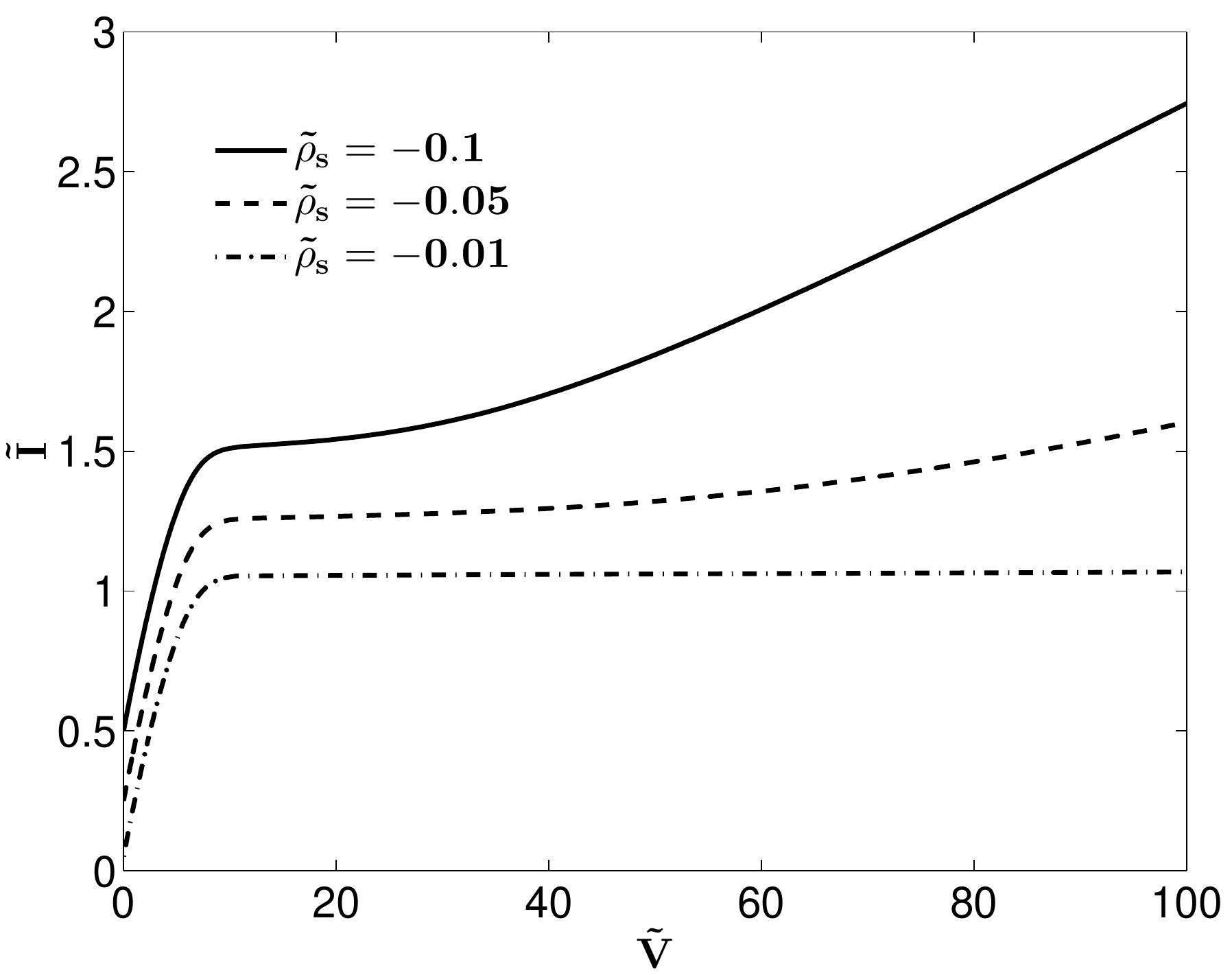}
\caption{Current-voltage relations for 1D model with normal flow (Fig.~\ref{fig:1dflow}) at a fixed, high flow rate, $Pe=5$, showing over-limiting current as  the surface charge is increased. The vertical shift of the curves is due to streaming current from the convection of  counter-ions screening the pore charge of the leaky membrane. }
\label{fig:1dflowiv}
\end{figure}

Figure~\ref{fig:1dflowiv} shows the current-voltage relationship for a case of strong flow, $\mbox{Pe}=5$. The shape of the curve is noticeable different from the case with no convection\cite{dydek2011overlimiting}, Eq. (\ref{eq:IV1D}), with a delayed, curved transition to the over-limiting regime. However, at higher voltages the overlimiting current eventually becomes linear, similar to the no-flow case.

\subsection*{ {\it Energy Cost of Deionization } } 

In the regime of over-limiting current, the flow-through electrode is continuously deionizing the fluid as it passes through the leaky membrane.   This setup could have applications to flow-through capacitive desalination~\cite{suss2012capacitive}, where the flow channel is filled with a porous medium or microchannel array, and it is a simple first approximation for the cross-flow geometry of shock electrodialysis~\cite{deng2013} discussed below.  For these applications, the model provides a simple case to analyze the energy cost of de-ionization.

The energy per volume of initial solution processed, $E_v$, is equal to input electrical power ($IV$) divided by the volumetric flow rate ($Q$):
\begin{equation}
E_v=\frac{IV}{Q}=2kTc_0\frac{\tilde{I}\tilde{V}}{\mbox{Pe}}
\end{equation}
which has the dimensionless form,
\begin{equation}
\tilde{E}_v=\frac{E_v}{2kTc_0}=\frac{\tilde{I}\tilde{V}}{\mbox{Pe}}.
\label{eq:1Denergy}
\end{equation}
The energy cost, $\tilde{E}_v$, is a function of the surface charge density, $\tilde{\rho}_s$, the applied current, $\tilde{I}$, and the velocity, $\mbox{Pe}$, each in  a suitable dimensionless form.

\begin{figure}[ht]
\centering
(a)\includegraphics[width=3in]{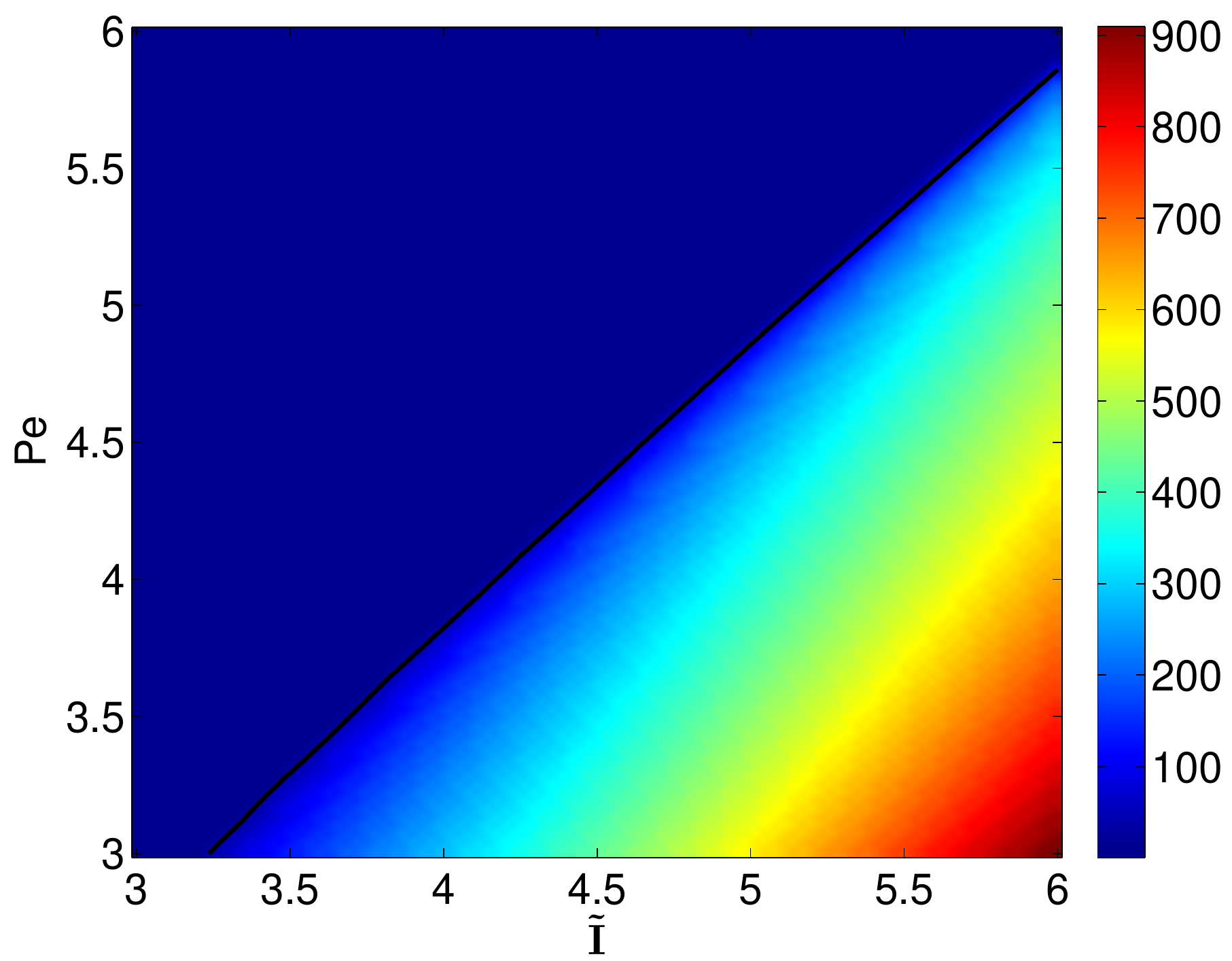}\\
(b)\includegraphics[width=3in]{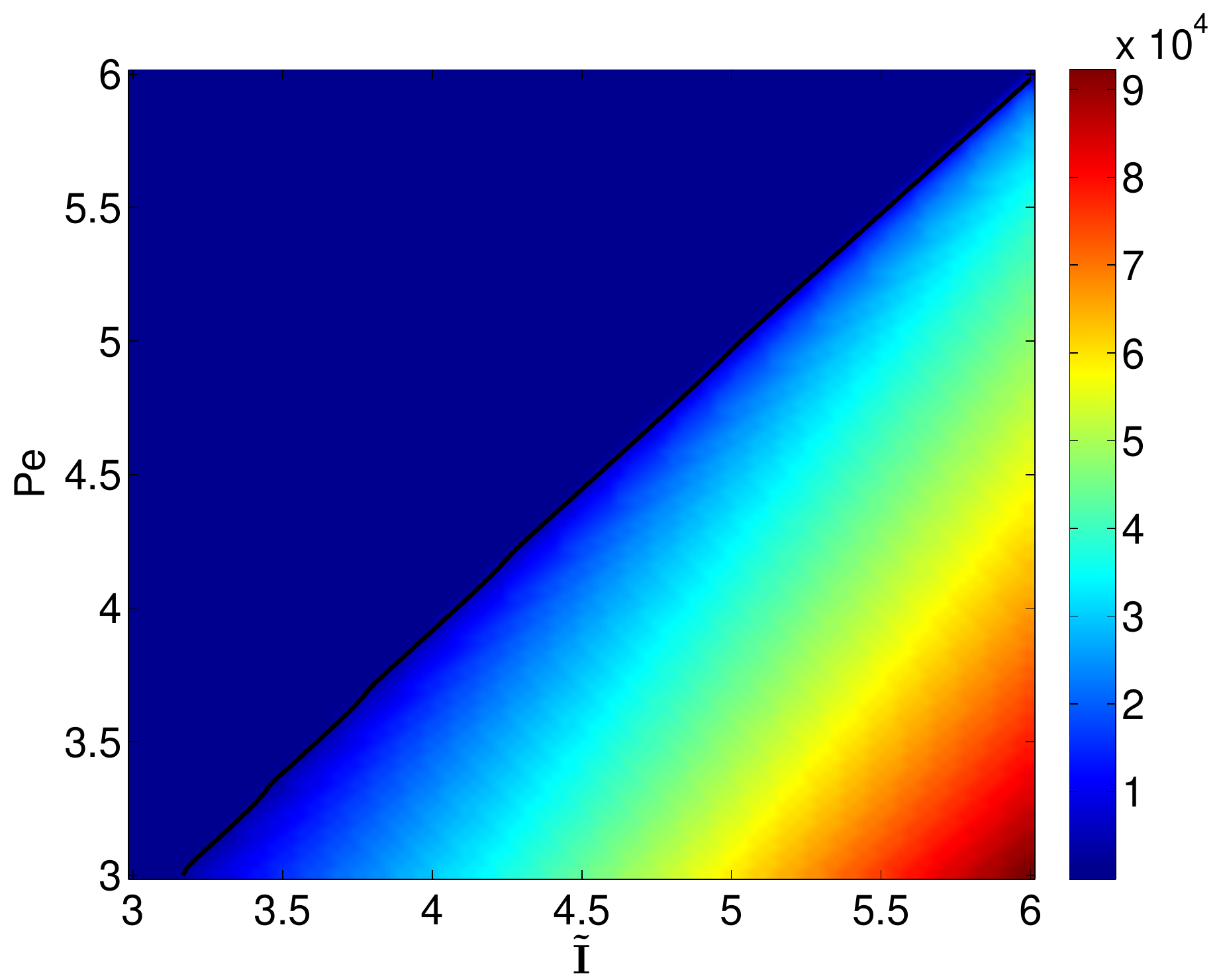}
\caption{Energy cost of deionization for the 1D model of normal flow through a leaky membrane and porous electrode. The black line indicates where a depletion region has formed. Below this line the outlet concentration, $\tilde{c}$, is 0.001 or less. (a) $\tilde{\rho}_s=-0.01$; (b) $\tilde{\rho}_s=-0.0001$}
\label{fig:1Denergy}
\end{figure}

Model predictions for a fixed system geometry are shown in Figure \ref{fig:1Denergy}. In these plots $\tilde{E}_v$ is shown versus varying $\mbox{Pe}$ and $\tilde{I}$ for two different values of $\tilde{\rho}_s$ (-0.01 and -0.0001). A black line indicating when the depletion region has formed ($\tilde{c}=0.001$) is placed on top of these surface plots. Below this line $\tilde{c}$ is less than 0.001 and the energy per volume increases. The energy profile appears very similar in these two plots, with the values differing by a factor of 100. In the 1D case, surface conduction goes as $\tilde{\rho}_s\tilde{V}$. Therefore as $\tilde{\rho}_s$ decreases by a factor of 100, the voltage must increase by a factor of 100 to maintain the same level of conduction. This increase in necessary voltage is what leads to the 100-fold increase in energy. 

In order to reduce the energy per volume required for deionization, the depletion region should be as small as possible. This is the reason that increasing the flow rate ($\mbox{Pe}$) decreases $\tilde{E}_v$. Similarly, increasing the applied current past the point of early depletion wastes energy and increases $\tilde{E}_v$. In this 1D model with uniform flow, the fluid recovery fraction, or ratio of deionized to incoming fluid volumes, is 100\%.  High fluid recovery is a hallmark of flow-through separations in porous media, but the shock phenomenon provides an opportunity for efficient separations in cross flow, perpendicular to the current, which we analyze next with a 2D model.

\section*{ Steady State with Cross Flow }

\subsection*{ {\it Fractionation by Deionization Shocks }}

The formation of a deionization shock represents a dynamic, ``membraneless" separation of salty and deionized solution, which can be exploited for water purification, brine concentration, or other separations by fractionation in cross flow.  In contrast to traditional electrodialysis (ED), in ``shock electrodialysis"~\cite{deng2013} there is no fixed physical, chemical or electrostatic barrier between the two regions that spontaneously form in a homogeneous porous medium. The strong localization of the salt concentration jump in the shock  and its ability to propagate to a desired position enables separation in cross flow.

\begin{figure}
\centering
\includegraphics[width=3.5in,keepaspectratio=true]{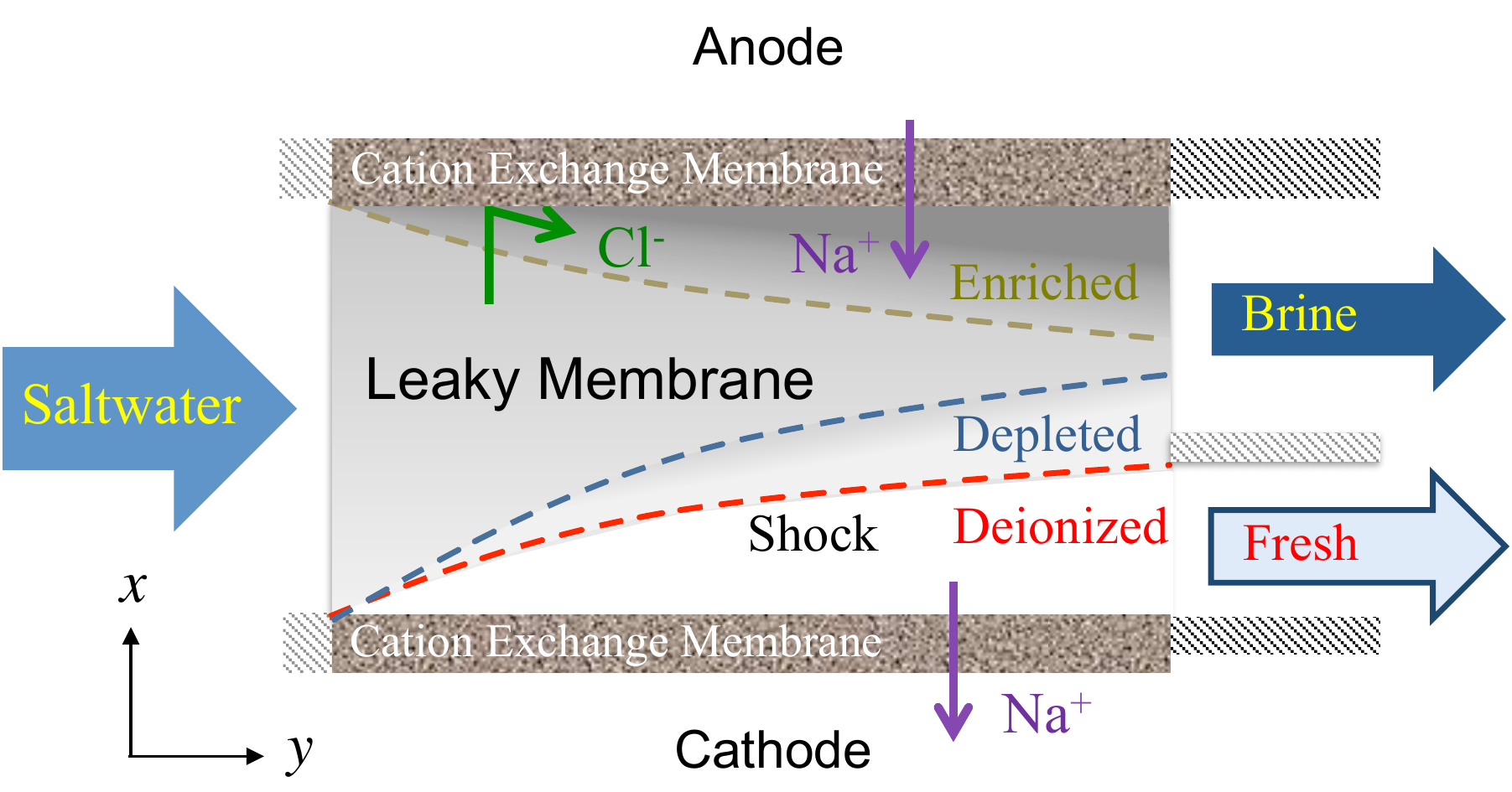}
\caption{  Sketch of one layer of a cross-flow shock electrodialysis system for water deionization and brine concentration. Current in the normal (vertical) direction passes though a negatively charged porous material (leaky membrane) sandwiched between two cation-exchange membranes (or other cation selective layers).  Water flows in the perpendicular direction and is split into brine and deionized streams upon exiting. }
\label{fig:schematic}
\end{figure}

The basic idea, sketched in Figure~\ref{fig:schematic}, is to drive fluid flow through the leaky membrane in one direction, while driving over-limiting current in the perpendicular direction between ion perm-selective membranes.  The deionization shock propagates in the cross flow to form a boundary layer of strong depletion, which extends across a fresh water collection outlet on the downstream end of the leaky membrane.  If the leaky membrane is sandwiched between identical ion-exchange membranes, then an enrichment diffusion layer also forms on the other side, which is collected in a brine stream, separated from the fresh stream by splitting the flow leaving the leaky membrane.   This layered structure is the basic building block of a scalable shock ED system with electrode streams on the ends to sustain the current.

Several parameters will affect the efficacy and the efficiency of such a device, including the surface charge of the leaky membrane, the geometry of the device, and the applied current or potential. In order to understand how these parameters relate to each other, we use a 2D LMM. Analytical solutions with nested boundary layer structure are possible in the relevant regime of fast cross flow, where the deionization shock and enrichment regions remain well separated~\cite{mani2013crossflow}, but here we focus on numerical solutions for finite geometries in a wide range of operating conditions.  The purpose of this model is to understand in a simple way how forced convection and SC affect ion transport while providing design guidelines to optimize the system.

\subsection*{ {\it Two Dimensional Model }}

Consider a leaky membrane of height, $h$, and length, $L$, where $x\in[0,h]$ and $y\in[0,L]$. Uniform flow
with velocity, $u$, in the $y$ direction originates from $y=0$. In this 2D model, we neglect axial diffusion, which is valid beyond a distance $D/u$ from the inlet, which is small if $\mbox{Pe}_y = uL/D \gg 1$.  In this regime,  convection dominates in the y-direction, and diffusion in the x-direction. The LMM conservation equations then take the form
\begin{equation}
\begin{split}
u\frac{\partial c_+}{\partial y}=D\left[\frac{\partial^2c_+}{\partial x^2}+\frac{ze}{k_BT}\frac{\partial}{\partial x}\left(c_+\frac{\partial\phi}{\partial x}\right)\right],\\ u\frac{\partial c_-}{\partial y}=D\left[\frac{\partial^2c_-}{\partial x^2}-\frac{ze}{k_BT}\frac{\partial}{\partial x}\left(c_-\frac{\partial\phi}{\partial x}\right)\right].
\end{split}
\end{equation}
Nondimensionalizing as before, with $\tilde{x}=\frac{x}{h}$ and $\tilde{y}=\frac{y}{L}$, the dimensionless conservation equations are
\begin{eqnarray}
\frac{uh^2}{L D}\frac{\partial\tilde{c}}{\partial\tilde{y}}= \frac{\partial^2\tilde{c}}{\partial\tilde{x}^2}-\tilde{\rho}_s\frac{\partial^2\tilde{\phi}}{\partial\tilde{x}^2} ,
\label{eq:concflow} \\
 \frac{\partial}{\partial\tilde{x}}\left[\left(\tilde{c}-\tilde{\rho}_s\right)\frac{\partial\tilde{\phi}}{\partial\tilde{x}}\right]=0.\hspace{3mm}
 \end{eqnarray}	

In our previous examples (Figs. ~\ref{fig:1D}, \ref{fig:1dflow}), the leaky membrane  was in contact with a reservoir of constant concentration at one end ($x=0$) and a cation-selective membrane or porous electrode at the other end ($x=h$). In this case, anions are blocked at both ends  ($x=0$ and $x=h$), which implies Neumann type conditions,
\begin{align}
&\tilde{x}=0: \tilde{\phi}=-\tilde{V},\ \frac{\partial\ln\tilde{c}}{\partial\tilde{x}}=\frac{\partial\tilde{\phi}}{\partial\tilde{x}}, \\
&\tilde{x}=1: \tilde{\phi}=0,\ \frac{\partial\ln\tilde{c}}{\partial\tilde{x}}=\frac{\partial\tilde{\phi}}{\partial\tilde{x}}, \\
&\tilde{y}=0: \tilde{c}=1.
\end{align}
with Dirichlet boundary conditions for the concentration at the inlet and potential at the membranes.  The current density (per area) is no longer uniform,
\begin{equation}
\tilde{J}(\tilde{y}) = -\left[\tilde{c}-\tilde{\rho}_s\right]\frac{d\tilde{\phi}}{d\tilde{x}},
\end{equation}
where $\tilde{J}=\frac{JL}{2zeDc_0}$ is the dimensionless current density in the x-direction. $\tilde{J}$ must be integrated over the membrane area to obtain the total current
\begin{equation}
\tilde{I} = \int_0^1 \tilde{J}(\tilde{y})d\tilde{y}
\end{equation}
Noticing that several parameters of interest are lumped together, the conservation equation can be rewritten in terms of the P\'eclet number
\begin{equation}
\mbox{Pe}=\frac{u h}{D}
\end{equation}
and a new axial length variable,
\begin{equation}
\hat{y}=\frac{y D}{u h^2}=\frac{L}{h}\frac{\tilde{y}}{\mbox{Pe}}
\label{eq:f}
\end{equation}
scaled to the entrance length for the convection-diffusion boundary layer, $u h^2/D$, as usual in the analysis of forced convection in a channel or pipe~\cite{deen_book}.
With this change of variables,
\begin{equation}
\frac{\partial\tilde{c}}{\partial \hat{y}}=\frac{\partial^2\tilde{c}}{\partial\tilde{x}^2}-\tilde{\rho}_s\frac{\partial^2\tilde{\phi}}{\partial\tilde{x}^2}.
\end{equation}
the conservation equation becomes the same as the 1D, transient equation, Eq.(~\ref{eq:conctrans}). Therefore, the solutions from the previous section can be reworked and applied here. 

\begin{figure}
\centering
\includegraphics[width=3in,keepaspectratio=true]{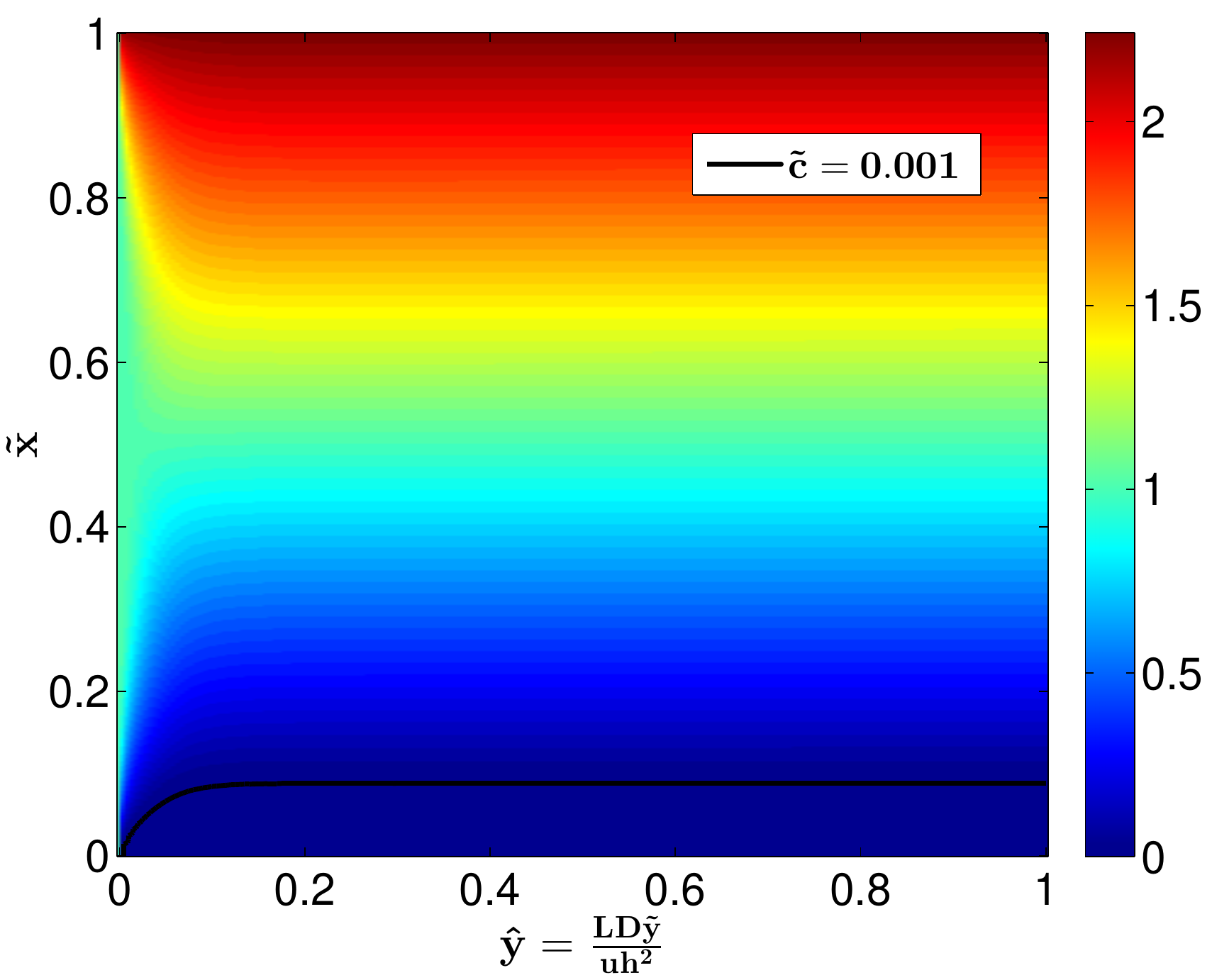}
\includegraphics[width=3in,keepaspectratio=true]{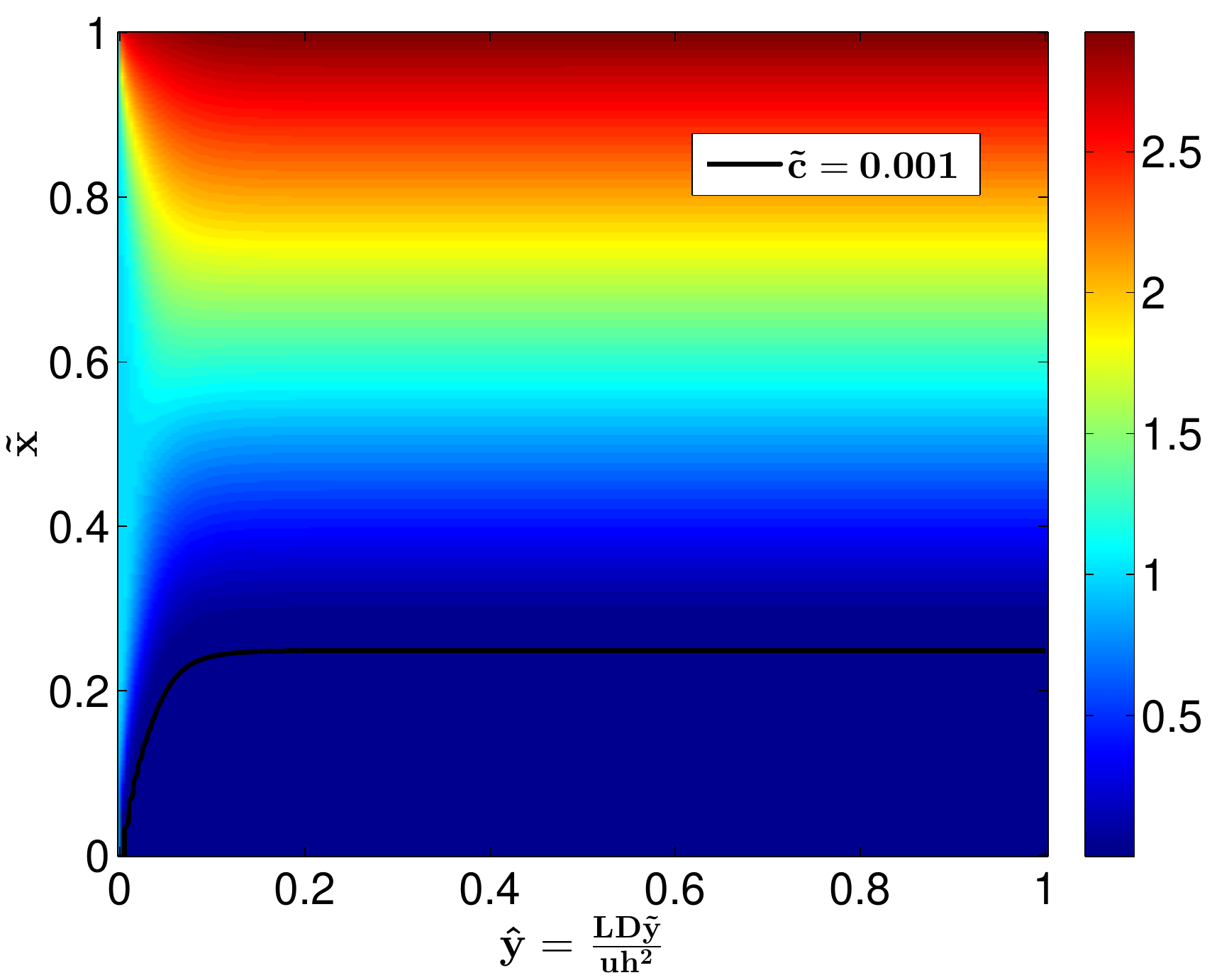}
\caption{ Steady concentration profile in a  simple 2D model of the shock electrodialysis device of Fig. ~\ref{fig:schematic} at high voltage $\tilde{V}=30$ for varying dimensionless surface charge in the leaky membrane:  (a) $\tilde{\rho}_s=-0.01$, b) $\tilde{\rho}_s=-0.05$.}
\label{fig:2Dflow}
\end{figure}

Example concentration profiles are shown in Figure~\ref{fig:2Dflow} for $\tilde{\rho}_s=-0.01,-0.05$ and $\tilde{V}=30$. As expected, increasing the surface charge increases the size of the depleted region, $\delta$. Here $\delta$ was taken to be the point where $\tilde{c}=0.001$. It is also important to note that at around $\hat{y}=0.1$ the concentration has reached its steady state value and no further depletion occurs. By setting $\tilde{y}=1$ these plots can be used to examine the outlet concentration distribution. For example, in the case of $\tilde{\rho}_s=-0.05$ (Figure~\ref{fig:2Dflow}b) the outlet can be fractionated at $\tilde{x}=0.25$ and if $\hat{y}>0.1$ only depleted fluid will be collected. This analysis can be used to determine the best geometry and flow rate. In order to maximize the flow rate, $\hat{y}$ should be minimized. In order to get full depletion, $\hat{y}$ should exceed the dimensionless distance to achieve steady state, which for these two cases is around 0.1; in other words, full depletion occurs at roughly 10\% of the entrance length. Additionally, scaling up the system will not be a linear process, since $\hat{y}\propto\frac{L}{h^2}$. 

\subsection*{ {\it Design Principles for Shock Electrodialysis } }

\begin{figure}
\centering
\includegraphics[width=3in,keepaspectratio=true]{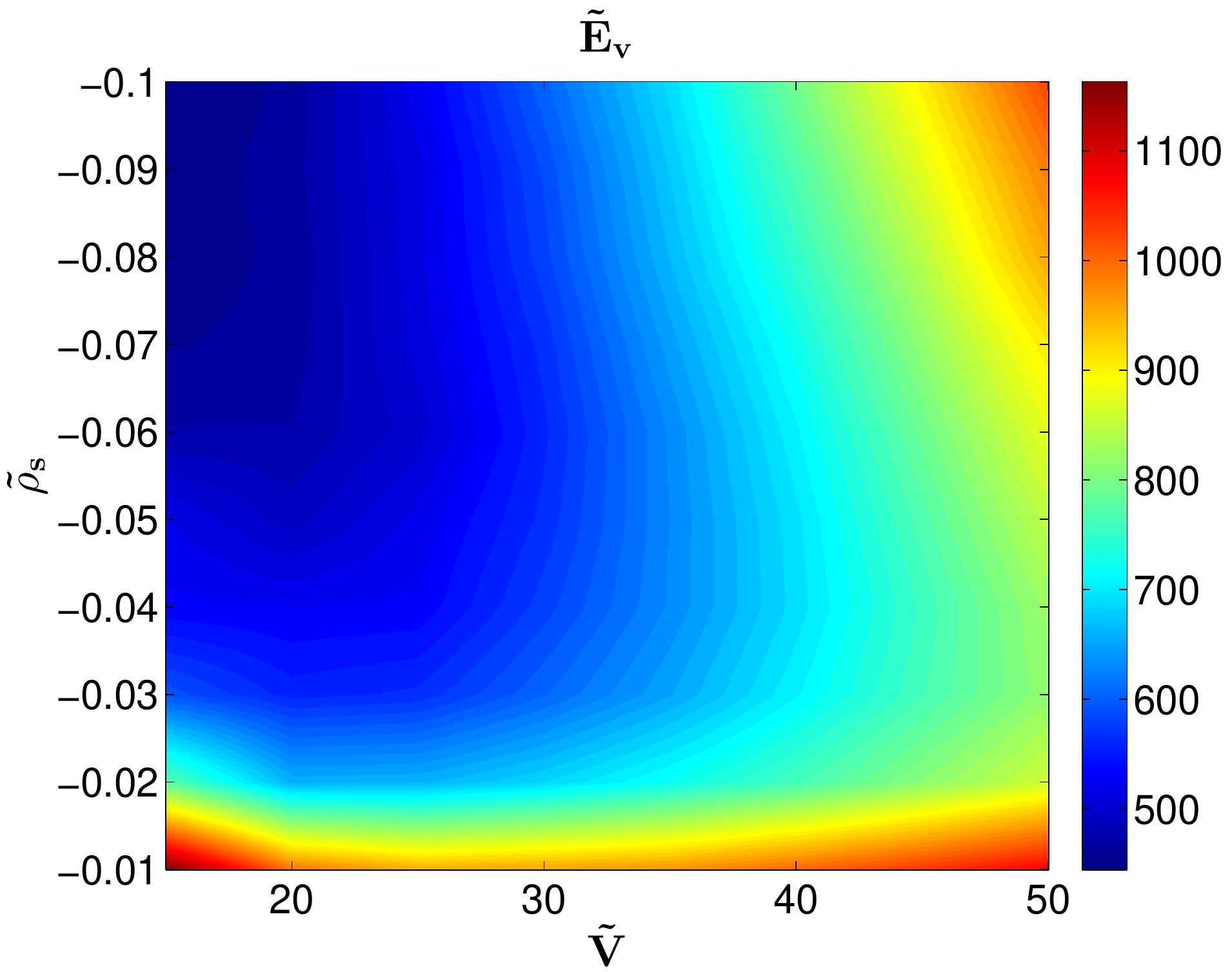}
\caption{Dimensionless energy $\tilde{E}_v$ (color contours) per volume of de-ionized fluid in the model of Fig.~\ref{fig:2Dflow}, versus the dimensionless surface charge $\tilde{\rho}_s$ and dimensionless voltage $\tilde{V}$, evaluated at $\hat{y}=0.1$. }
\label{fig:energy}
\end{figure}

An important characteristic of this extraction technique is the energy required to create the necessary depletion region. For a volumetric flow rate of $Q$, with current density, $I$, and voltage, $V$, the energy $E_v$ required per volume of depleted solution is given by 
\begin{equation}
E_v=\frac{IV}{\delta Q}=\frac{2k_BTc_0}{\mbox{Pe}}\frac{\tilde{I}\tilde{V}}{\delta},
\label{eq:energy}
\end{equation}
The parameters $\tilde{I}$, $\tilde{V}$, and $\delta$ are related through the conservation and current equations. If $\hat{y}$ is far enough downstream to reach steady state, then $\tilde{V}$ and $\delta$ can be found based solely on $\tilde{I}$ and $\tilde{\rho}_s$. As a result, it is useful to consider the dimensionless energy efficiency, 
\begin{equation}
\tilde{E}_v=\frac{E_v\mbox{Pe}}{2k_BTc_0}=\frac{\tilde{I}\tilde{V}}{\delta}
\end{equation}
A plot of $\tilde{E}_v$ over a range of $\tilde{V}$ and $\tilde{\rho}_s$ values is shown in Figure~\ref{fig:energy}. This plot was generated for a system with a maximum length of $\hat{y}=0.1$, corresponding to about 10\% of the entrance length. At lower to moderate applied voltages, increases in the surface charge density lead to decreases in the depletion energy. While increases in $|\tilde{\rho}_s|$ will lead to increases in $\tilde{I}$ the corresponding increases in $\delta$ are sufficient to lower the required energy. However, at higher applied voltages the balance is shifted and the increase in power cost overwhelms the efficiency gained by creating a larger depletion region. Once $\tilde{V}$ and $\tilde{\rho}_s$ have been determined, the energy efficiency can be calculated. This efficiency can be enhanced by properly designing the system geometry. For example, the larger the aspect ratio, $L/h$, the lower the energy efficiency, as seen in Eq. (\ref{eq:energy}). 

\begin{figure}
\centering
\includegraphics[width=3in,keepaspectratio=true]{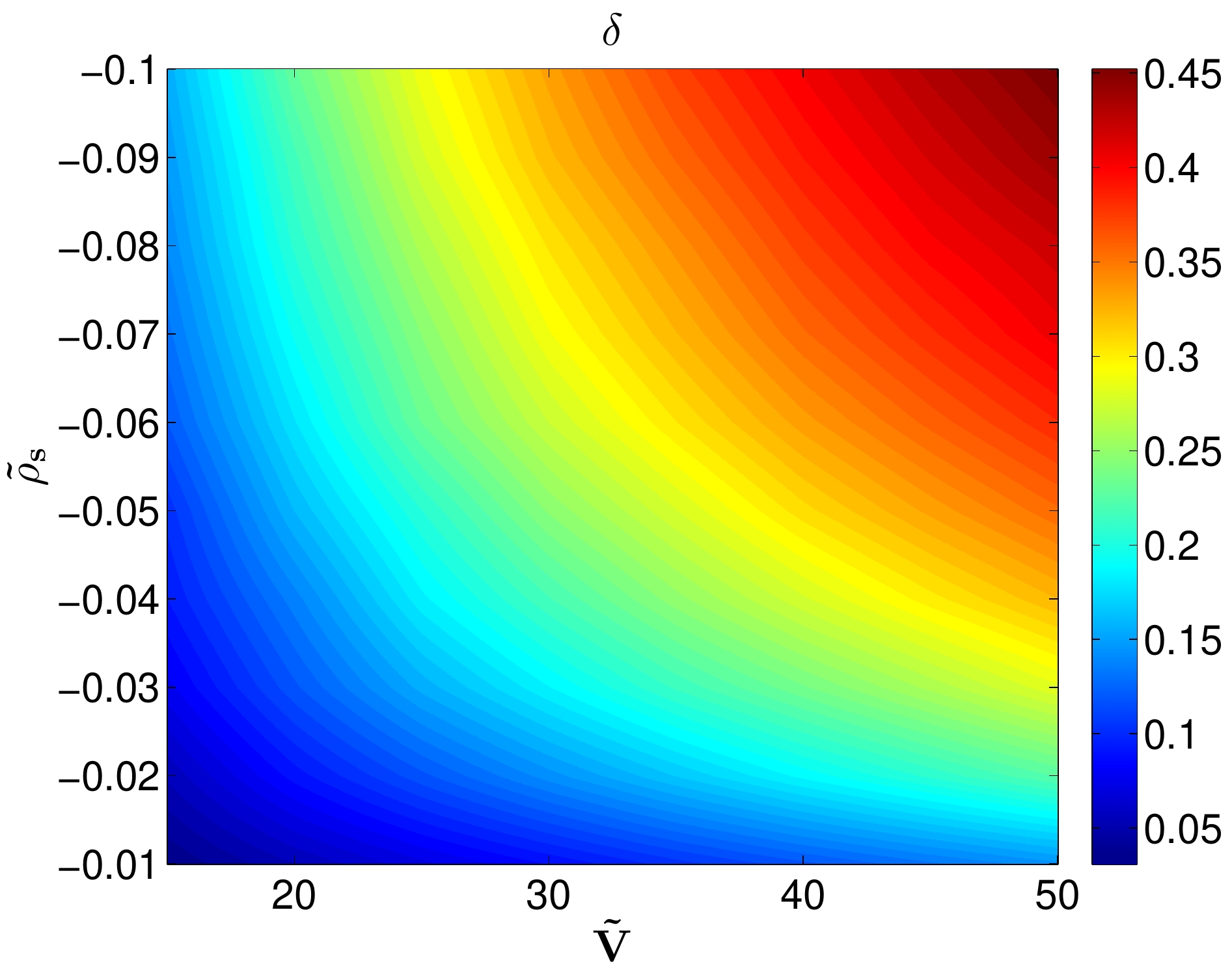}
\caption{Water recovery $\delta$ (ratio of deionized to incoming fluid volumes) in the model of Fig.~\ref{fig:2Dflow}, which increases with increasing $\tilde{V}$ and $\tilde{\rho}_s$.}
\label{fig:delta}
\end{figure}

In addition to energy requirements, in order to develop a practical device, the volume of depleted fluid relative to the incoming fluid (the ``water recovery" percentage in desalination) must be considered. A very efficient device that only depletes 1\% of an incoming stream may not be particularly desirable. In this case many passes would be required to achieve a sufficient amount of depleted solution. Recovery in this model corresponds to the size of the depletion region, $\delta$. A plot of $\delta$ versus $\tilde{V}$ and $\tilde{\rho}_s$ is shown in Figure \ref{fig:delta}, under the same conditions as in Figure \ref{fig:energy}. The region of highest recovery does not correspond to the region of lowest energy per volume (Figure~\ref{fig:energy}). Therefore a balance must be struck between the two values, depending on the requirements of a desired system. 

In order to fully design a practical device that utilized SC in flow, one other parameter needs to be addressed. Throughout this study the parameter $\tilde{\rho}_s$ has played an important role. However, this parameter is a function of the initial anion concentration, $c_0$. The power of SC goes up as $c_0$ goes down. In order to have an effect on higher ion concentrations, the volume surface charge density, $\rho_s$ of the material should be increased. This can be done by either altering the surface charge of the material, $\sigma_s$ or decreasing the pore size. For instance, a typical silica bead in water has a surface charge density of about -0.001 $\mathrm{coul/m^2}$ \cite{behrens2001charge}. In a 1mM solution, a porous structure of these beads with a pore size of 10 $\mu$m will result in $\tilde{\rho}_s=-0.001$. However, if the pore size decreases to 100 nm, then $\tilde{\rho}_s=-0.1$ and SC plays a more dominant role. Based on this analysis, the smaller the pores, the better the de-ionization. However, the energy analysis conducted here did not take into consideration the force to pump the fluid through the porous material. As the pore size decreases the pump energy required increases. As a result, decreasing the pore size may not be the best solution. Alternatively, the surface of the porous material can be altered to create a more negative surface. In order to maximize the energy efficiency, the device should be designed with a high aspect ratio. Additionally, the velocity should be maximized such that $\hat{y}$ defined in Eq.~(\ref{eq:f}) is kept low but at a steady state value. Based on Figures~\ref{fig:energy} and~\ref{fig:delta}, the applied current should be above the limiting value but low enough such that the energy per volume and water recovery are at acceptable levels. In this manner, an efficient SC-flow device can be developed using simple materials. 

\section*{ Conclusion }

Unlike ideal ion-exchange membranes, which maintain a large conductivity of counter-ions, the conductivity of ``leaky membranes" with larger pores and/or smaller surface charge densities can vary significantly in response to a large applied voltage. The surface conductivity, which remains even if the bulk salt is depleted, provides a mechanism for over-limiting current, faster than diffusion. This can lead to a macroscopic region of salt depletion behind a propagating deionization shock, which opens new possibilities for nonlinear electrokinetic separations in porous media.   Building on recent work~\cite{mani2009propagation,mani2011desalination,yaroshchuk2012acis}, we formulate a general Leaky Membrane Model and derive representative analytical and numerical solutions for finite domains.  We focus on the simplest situation of a symmetric binary electrolyte in a leaky membrane of constant surface charge density, uniform pressure-driven flow, negligible hydrodynamic dispersion, and no electro-osmotic flow. 

Relaxing these assumptions and deriving suitable modifications of the model provide challenging avenues for research.  For example, charge regulation in a multicomponent electrolyte due to specific adsorption of ions is a classical source of nonlinearity~\cite{helfferich_book,rhee_book}, which in leaky membrane can lead to over-limiting current by ``current-induced membrane discharge"~\cite{andersen2012}. The LMM with charge regulation could have relevance for electrokinetic remediation in soils\cite{probstein1993,shapiro1993,jacobs1994,kamran2012}, as well as ion transport in biological cells.  The dynamics of charged colloids in leaky membranes may also lead to interesting nonlinear dynamics, generalizing shock waves in capillary electrophoresis~\cite{ghosal2010,ghosal2012,chen2012a,chen2012b,chen2012c}. The LMM may also improve the accuracy of porous electrode theories, which currently assume electroneutrality in the solution phase and neglect surface conduction~\cite{newman_book}, which already account for capacitive charging of double layers~\cite{biesheuvel2010} with Faradaic reactions~\cite{biesheuvel2011,biesheuvel2012} and specific adsorption of intercalation reactions~\cite{ferguson2012,ACR2013}, but generally neglect surface conduction.  In all of these situations, perhaps the most difficult and important extension of the LMM will be to account for electro-osmotic flow and associated dispersion phenomena at the macroscopic scale~\cite{yaroshchuk2011coupled,dydek2011overlimiting}.

\bibliographystyle{aichej} 
\bibliography{scbib}

\end{document}